# AM-DefectNet: Additive Manufacturing Defect Classification Using Machine Learning - A comparative Study

**Mohsen Asghari Ilani[1], Yaser Mike Banad[1]**

[1] School of Electrical and Computer Engineering, University of Oklahoma, Norman, 73019, U.S.A.

## Abstract

Additive Manufacturing (AM) processes present challenges in monitoring and controlling material properties and process parameters, affecting production quality and defect detection. Machine Learning (ML) techniques offer a promising solution for addressing these challenges. In this study, we introduce a comprehensive framework, AM-DefectNet, for benchmarking ML models in melt pool characterization, a critical aspect of AM. We evaluate 15 ML models across 10 metrics using 1514 training and 505 test datasets. Our benchmarking reveals that non-linear tree-based algorithms, particularly CatBoost, LGBM, and XGBoost, outperform other models, achieving accuracies of 92.47%, 91.08%, and 90.89%, respectively. Notably, the Deep Neural Network (DNN) also demonstrates competitive performance with an accuracy of 88.55%. CatBoost emerges as the top-performing algorithm, exhibiting superior performance in precision, recall, F1-score, and overall accuracy for defect classification tasks. Learning curves provide insights into model performance and data requirements, indicating potential areas for improvement. Our study highlights the effectiveness of ML models in melt pool characterization and defect detection, laying the groundwork for process optimization in AM.

**Keywords**: Additive Manufacturing, Machine Learning, Learning Curve, Melt pool.

## 1. Introduction

Transitioning from conventional manufacturing, which relies on physical-contact energy to shape materials, to advanced manufacturing driven by non-contact energy holds promise for meeting the diverse demands of various industries such as biomedical, electronics, and aerospace applications. Additive Manufacturing (AM), commonly known as 3-D printing, stands as a pioneering and disruptive technology at the forefront of the next industrial revolution, facilitating the production of increasingly complex geometries. In contrast to conventional manufacturing methods like milling, AM offers numerous advantages, including reduced material wastage, shorter lead times, and enhanced performance, reliability, and accuracy, particularly in the fabrication of customized designs with intricate geometries. These unique characteristics have propelled the commercialization of AM across high-tech sectors such as aerospace, automotive, energy, and biomedical fields. Despite garnering significant attention from academia and industry, challenges persist regarding the impact of direct laser energy and thermal energy on the material properties and surface integrity of AM-built products. Of particular concern is the quality of AM-built parts, as defects can undermine their structural integrity. Overcoming these challenges is daunting due to the complex, multi-scale physics inherent in the AM process

To gain a thorough understanding of how defects are generated and their impact on surface integrity and material properties, two monitoring approaches are commonly used: in-situ and ex-situ methodologies. In-situ monitoring involves controlling process parameters during the manufacturing process itself, while ex-situ monitoring provides an overview of the process output by analyzing changes in process parameters before and after machining. Common in-situ monitoring techniques include signal processing using acoustic sensors and thermography of the melt pool. Ex-situ monitoring techniques often involve image processing based on X-ray and computed tomography (CT) scans, along with numerical or categorical datasets. These techniques are crucial for detecting and controlling defect formation, such as lack of fusion, keyhole porosity, and balling, which are often linked to the dynamics of the melt pool. The characteristics of the melt pool, including its depth, width, and length, as well as key AM process parameters like laser power, scanning speed, hatch spacing, layer thickness, and beam diameter, strongly influence the defect mechanism in AM. Therefore, precise control and monitoring of the manufacturing process are essential for achieving high accuracy, repeatability, and low defect rates, resulting in production with desired properties.

The complexity of AM processes stems from several factors, including challenges in controlling melt pool geometry, thermal effects on melt flow, Marangoni and buoyancy forces, and the Heat-Affected Zone (HAZ). These processes involve changes in material properties and microstructure during various phases like heating, melting, evaporating, and cooling. This multi-physics and multi-scale nature of AM processes, combined with the significant influence of processing parameters on the quality of printed products, has prompted a shift in AM approaches. Traditionally, AM relied solely on physics-based methodologies. However, there has been a transition towards combining physics-based and data-driven approaches, as the application of machine learning holds increasing promise for predicting more accurate and reliable models in a cost-effective and time-efficient manner. Datasets extracted primarily focus on melt pool characteristics, given their critical role in defect formation in AM-built products. Consequently, learning (ML) algorithm based on input features correlated with targets is essential to accurately capturing input-target relationships. The multi-physics nature of AM processes introduces numerous factors influencing defect formation, such as melt pool geometry, microstructure, material properties, and melt pool dynamics during heating, melting, and cooling phases. This complexity extends to identifying linear, non-linear, or combined relationships among these factors. Consequently, creating ML models for AM faces challenges due to limited data availability and the intricate nature of AM processes.

Despite these obstacles, researchers have implemented ML techniques in laser powder bed fusion (LPBF) [1–6]. In ML, classification aims to forecast labels linked with given datasets, ranging from classes, categories, to targets. This process involves constructing a mapping function (f) from input variables (X) to discrete output variables (y), often categorized as binary or multi-class. Among the arsenal of ML tools, the Support Vector Machine (SVM) stands out, proficient in both classification and regression tasks. Scholars such as Khanzadeh et al. [7] extensively utilized SVM alongside other methods like Decision Trees (DT), K-Nearest Neighbors (KNN), Linear Discriminant Analysis (LDA), and Quadratic Discriminant Analysis (QDA) to detect defects, achieving notable accuracies (such as SVC with linear kernel:90.7%, SVC with polynominal data-driven analysis and machine learning (ML) have become integral to advanced manufacturing applications, particularly in AM research. While experimental in-situ monitoring techniques are effective, they can be costly, inefficient, and require extensive preparation and calibration. Alternatively, employing ML models built on experimental data offers a more cost-effective solution. With a reliable training dataset, ML models can make accurate predictions and efficiently determine optimal processing parameters, benefiting future AM setups.

Adapting laser-based and heat-affected AM processes to different machine setups, tweaking process parameters, laser types, and material properties, alongside maintaining controlled environments, poses a significant challenge in developing reliable datasets or models for widespread use. Additionally, gathering specific data for AM operations is both time-consuming and costly. Despite relying on calibrated sensors for accurate measurements through in-situ and ex-situ monitoring, limitations persist in data availability for AM processes. Selecting the right machine

kernel:97.97%, 98.44% with KNN, 90.7% for DT, 71.15% for LDA, and 98.21% for QDA). for predicting porosity. Additionally, SVM was applied by Scime et al. [8] and Gobert et al. [9] specifically for defect detection in additive manufacturing, achieving over 80% accuracy in identifying various defects. Moreover, Bayesian classifiers and Artificial Neural Networks (ANNs) have found roles in defect detection, with Bayesian classifiers offering probabilistic defect information in processes like Laser Beam Additive Manufacturing (LBAM), and ANNs demonstrating high accuracy when trained with labeled datasets. Researchers like Tapia et al. [10], Scime et al. [8], Lee et al.[11], Yuan et al. [12], and Gaikwad et al.[13] have contributed significantly to ML's advancement in AM, developing predictive models and frameworks to improve part quality and process optimization. These efforts have yielded substantial progress in forecasting parameters like melt pool depth, geometry, width, and more importantly AM defect as LOF, balling, and porosity.

In this study, our objective is to analyze melt pool behavior, particularly the types of defects, using a set of machine learning methods designed specifically for additive manufacturing. Referred to as AM-DefectNet, these methods utilize an extensive experimental dataset compiled by Akbari et al. [1], which includes data from various processing parameters, materials, and types of additive manufacturing processes (such as Powder Bed

Fusion and Directed Energy Deposition). Our approach involves constructing 16 machine learning models within AM-DefectNet, aiming to optimize processing parameters for identifying desired melt pool characteristics and common defects like lack of fusion (LOF), Balling phenomena, and keyhole formation in additive manufacturing parts. Additionally, we explore how different parameters associated with the manufacturing process impact the performance of these models. Furthermore, we introduce a data-driven Deep Learning (DL) model classification method, which offers greater interpretability compared to traditional machine learning models. This DL model aims to uncover complex relationships within the dataset, including additive manufacturing process types, processing parameters, material properties, and melt pool geometries, to enhance defect classification. We will compare the performance of this DL method with that of the machine learning models to assess its effectiveness in defect classification within additive manufacturing processes.

## 2. Methodology

In ***Figure 1***, the AM-DefectNet framework is depicted, encompassing the raw dataset features, the process of featurization, the employed ML models, and the target classification. This section delves into the processes of dataset collection and curation, feature engineering, and selection of ML algorithms.

### *2.1 Data Collection*

The data concerning both melt pool geometry and flaw and their types were gathered from peer-reviewed publications in manufacturing and materials journals. Particularly, emphasis was placed on studies presenting experimental data relevant to these characteristics, with a primary reference being the research conducted by Akbari et al. [1]. Additionally, information on the processing parameters and material properties employed in each experiment was compiled. These details are intended to be utilized as input variables for our ML models.

### *2.2 Datasets*

AM-DefectNet is constructed using data gathered from literature sources. Our existing dataset comprises approximately 2019 data points. Each data point includes information on processing parameters and material properties, as well as the geometry of the melt pool (Width, Length, and Depth) serving as input features. Additionally, the dataset includes labels indicating the type of defects present, such as Lack of Fusion (LOF), Balling, Desirable, and Keyhole.

As previously mentioned, our study considers various parameters that influence melt pool geometry, characteristics, and physical properties as inputs. These parameters include beam power, scanning speed, layer thickness, depth of the melt pool, specific energy in each laser spot, scanning pattern, laser properties, and material thermal and physical properties. A melt pool with a semi-circular shape and devoid of any defects or porosities is referred to as a desirable melt pool. However, in cases of high energy density, where the melt pool depth exceeds half of the melt pool width, the assumption of a semi-circular melt pool shape no longer holds, and the melt pool can be classified as being in keyhole mode. Keyhole porosity is observed in regions characterized by high power and low velocity, while lack-of-fusion voids occur in regions with low power and high velocity. Additionally, lack-of-fusion defects primarily occur due to insufficient overlap between adjacent melt pools or layers, often resulting from inadequate energy input or excessively large hatch distances. Balling phenomena also arise in regions with high power and velocity.

In order to gain a deeper understanding of the primary sources of defect formation, additional information was collected on material composition, alloys, AM process techniques, and the AM process parameters that impact process performance and melt pool characteristics. This comprehensive data collection aimed to provide insights into the intricate details of AM and the reasons behind defect formation. Subsequently, machine learning algorithms were employed using the aforementioned features to predict various properties of the melt pool. These properties were predicted using different ML classification models to determine the defect mode of the melt pool. Among the 16 models utilized, conventional ML models such as 'XGBoost', 'LGBM', 'AdaBoost', 'LogisticRegression', 'DecisionTree', 'RandomForest', 'CatBoost', 'k-NN', 'Voting', and 'Bagging' were deployed. Additionally, models including 'DNN', 'SVC with RBF kernel', 'SVC with linear kernel', 'SVC with polynomial kernel', 'SVC with sigmoid kernel', and Neural Networks (NNs) as Multilayer Perceptron (MLP) were employed for a comprehensive comparison of ML models in defect classification, a comparison not previously explored in this manner.

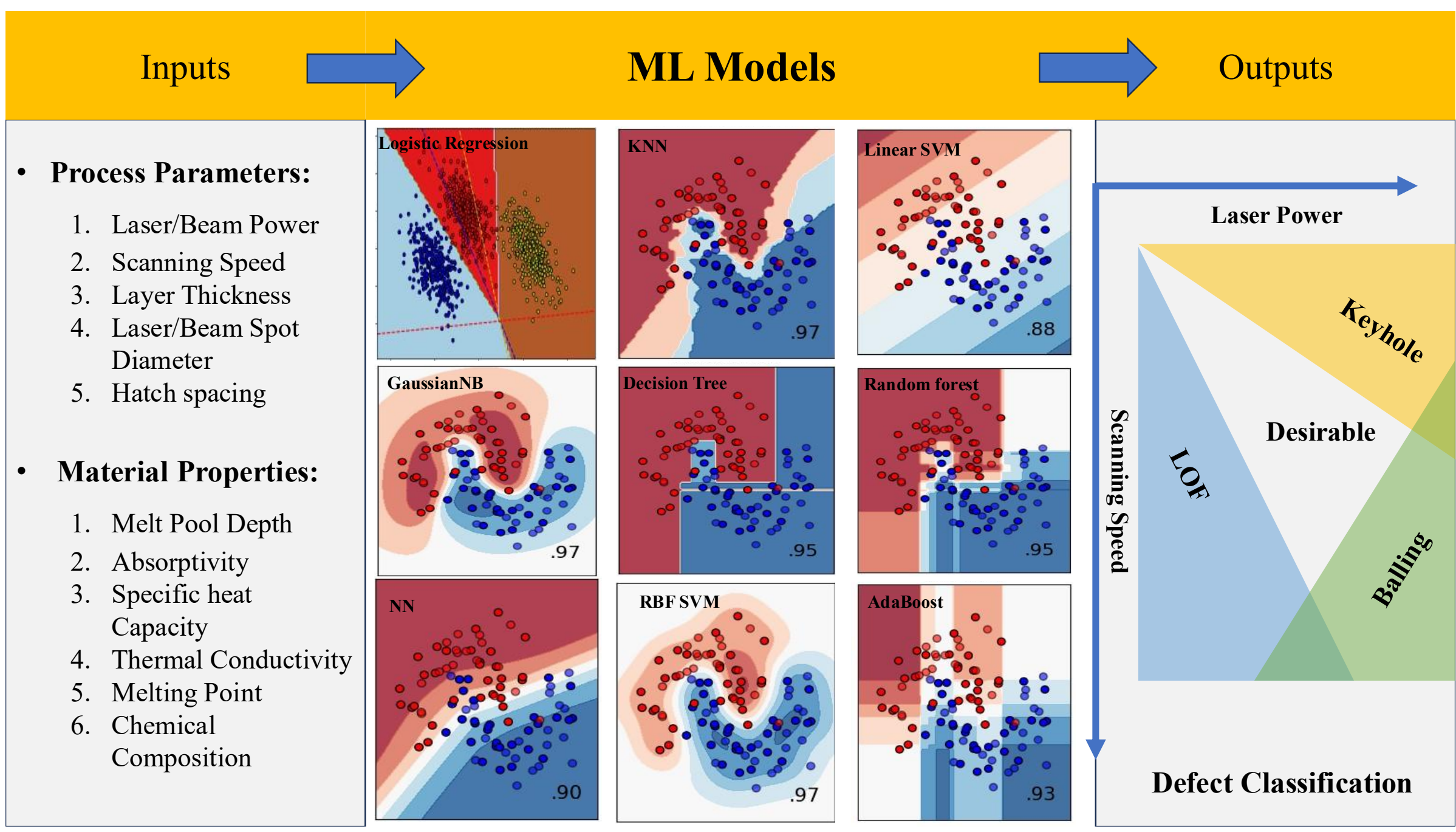

**Figure 1.** Inputs, ML models, and Outputs along with the task implemented in our AM-DefectNet benchmark.

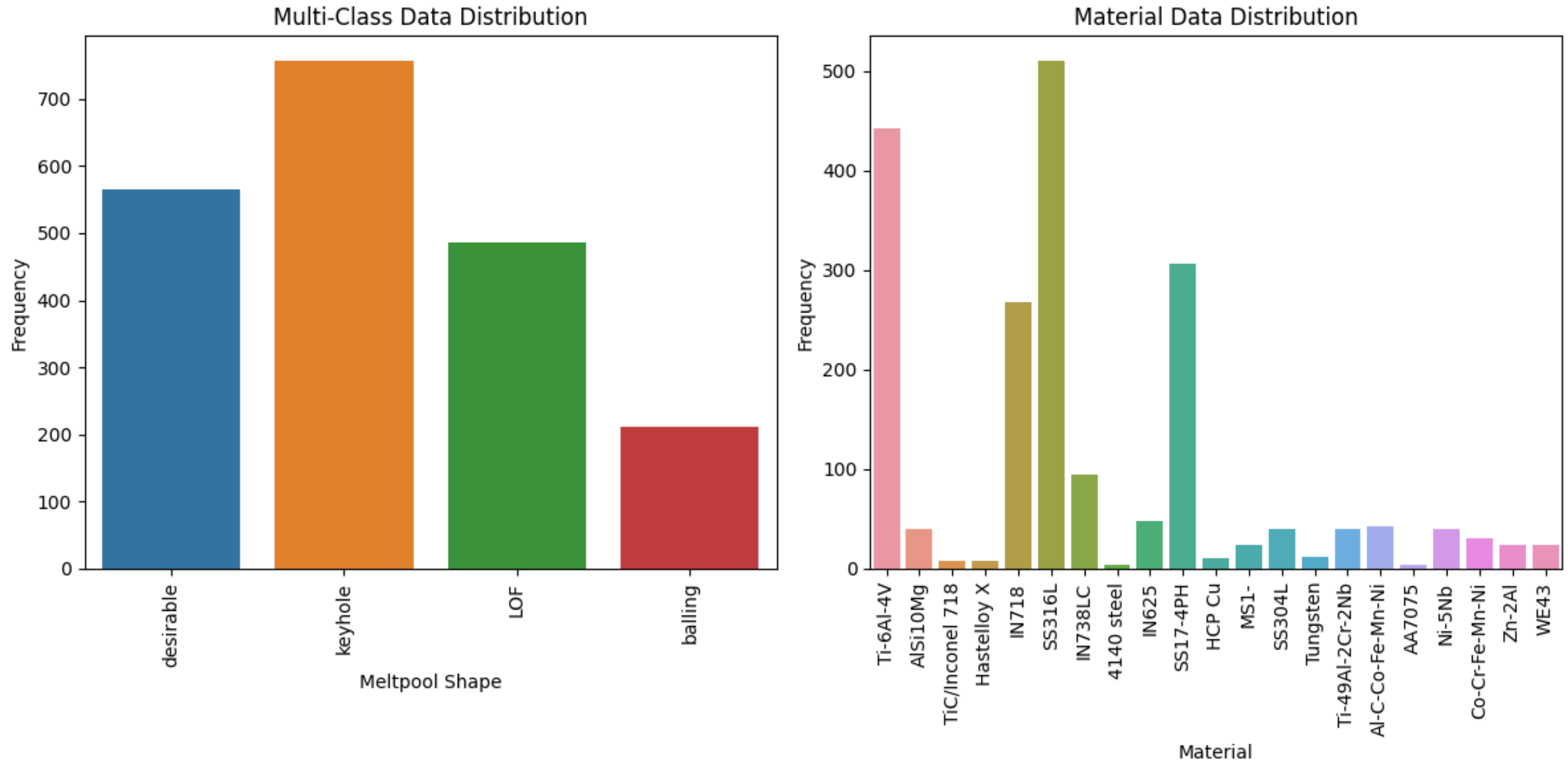

**Figure 2.** Distribution of Defect **(a)** and Material **(b)** classifications in our datasets.

Our aim is to categorize defects in AM by considering AM process parameters and material properties. ***Figure 2*** displays the distribution of defects (***Figure 2***a) categorized into four common classes observed during AM build processes: desirable (no defect present), keyhole, lack of fusion, and balling, along with the alloys used in the materials (***Figure 2***b) across the AM process, encompassing 21 commonly studied alloys. Our datasets indicate a higher prevalence of keyhole defects, notably in materials such as SS316L, Ti-6Al-4V, and SS17-4PH, frequently utilized by researchers and industry practitioners.

To further explore these details and effectively illustrate the characteristics associated with these four classes, we have integrated two commonly utilized additive manufacturing techniques: Selective Laser Melting (SLM) and Electron Beam Melting (EBM) into the datasets. As depicted in ***Figure 3***, the defect classification based on melt pool shape (***Figure 3a***) reveals a higher probability of keyhole defects occurring in SLM, while lack of fusion (LOF) defects are more prevalent in EBM. Additionally, ***Figure 3b*** showcases the alloys used in AM under the techniques of SLM and EBM, with a greater tendency towards using stainless steel and Ti alloys in SLM, whereas the opposite trend is observed in EBM.

Moreover, the distributions of laser/beam power, scanning speed, melting point, beam diameter, depth of melt pool, and thermal conductivity in Powder Bed Fusion (PBF) and Electron Beam Melting (EBM) processes for melt pool shape classification and alloys, respectively, as studied in our benchmark, are depicted in ***Figure 4*** and ***Figure 5***. As illustrated in ***Figure 4a***, a higher beam powder in EBM resulted in more occurrences of balling and desirable defects, whereas in SLM, more laser powder led to a higher likelihood of balling. Conversely, in ***Figure 4b***, a higher beam power was utilized in EBM of IN718, while SLM employed . HCP Cu. Regarding scanning speed, balling and lack of fusion were more likely to occur for both EBM and SLM, as shown in ***Figure 4c***, while stainless steel alloys were commonly used in high-speed SLM, as depicted in ***Figure 4d***. In ***Figure 4e***, a higher melting point of powders in both EBM and SLM increased the probability of four common defects in additive manufacturing. Alloys with higher melting point properties, such as Tungsten in SLM and IN718 for EBM, were found to exhibit this effect, as shown in ***Figure 4f***.

For further examination of parameters affecting melting geometry, ***Figure 5(a-f)*** demonstrate the impact of the Heat Affected Zone (HAZ) of laser spot diameter (***Figure 5 a, b***), depth of the melting pool (***Figure 5c, d***), and thermal conductivity (***Figure 5e***) on defect formation. These parameters also affect the likelihood of phase changes in crystal and material structure due to defects in the AM process. In ***Figure 5f***, it is noted that high thermal conductivity enables efficient heat transfer within the material during the melting and solidification stages, contributing to uniform heating and cooling rates. However, its effect on alloys with higher ranges of thermal conductivity was observed to be less significant

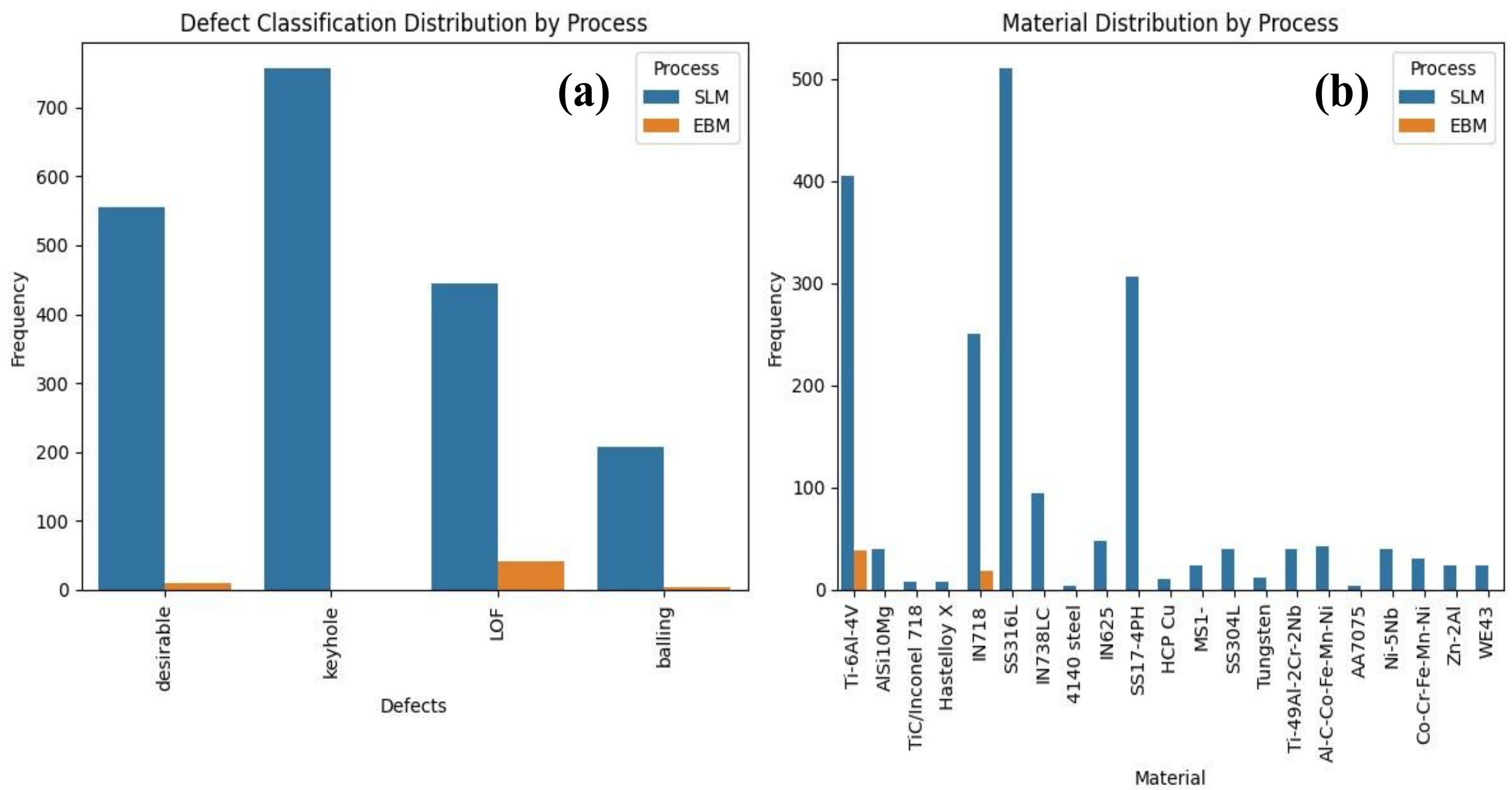

**Figure 3.** Distribution of Defect **(a)** and Material **(b)** classifications by sub-categorical AM processes as SLM and EBM in our datasets.

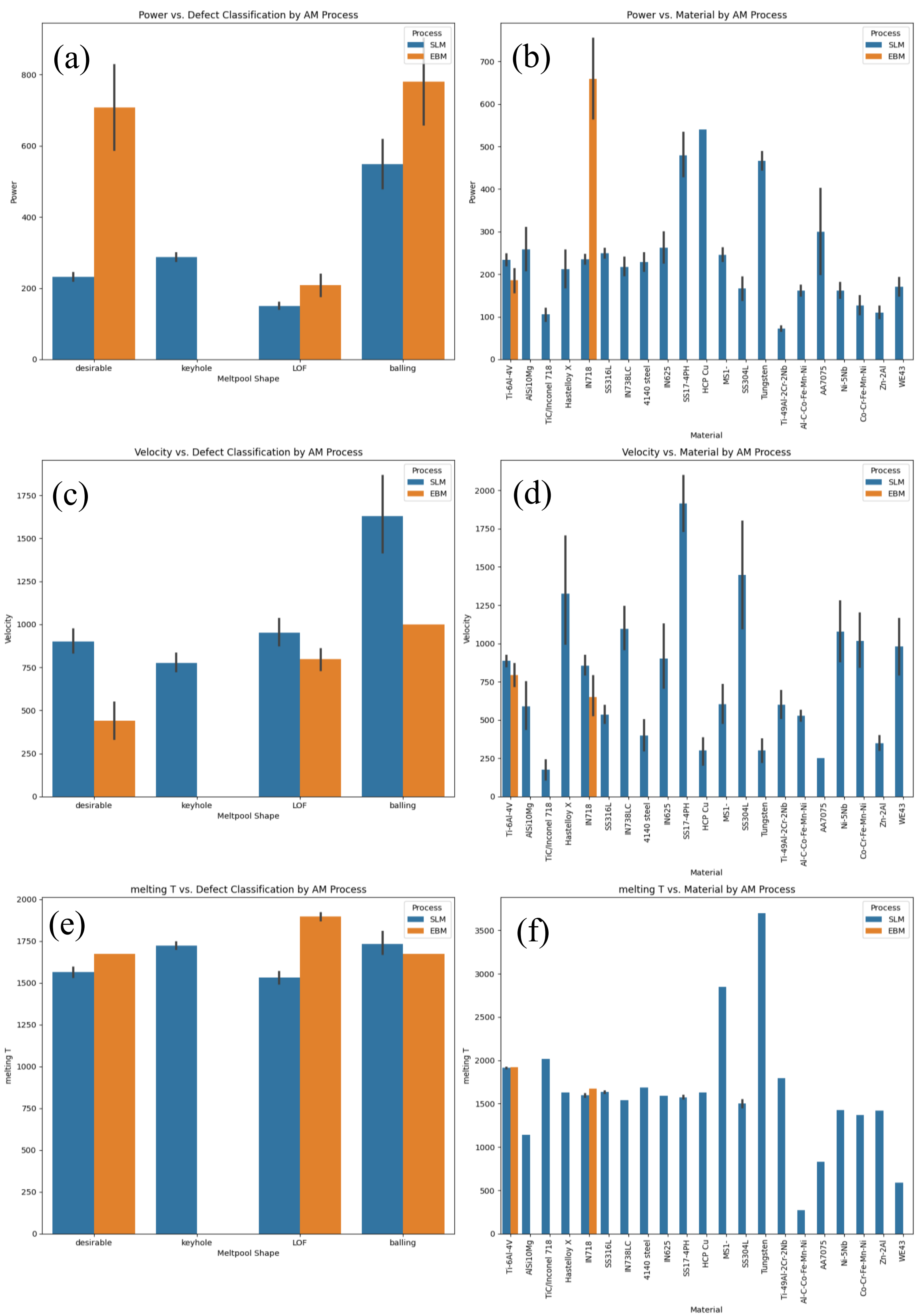

**Figure 4.** Distribution of Defect and Material classifications by sub-categorical AM processes as SLM and EBM in our datasets under **(a,b)** Laser Power, **(c,d)** Scanning speed, and **(e,f)** Melting Point.

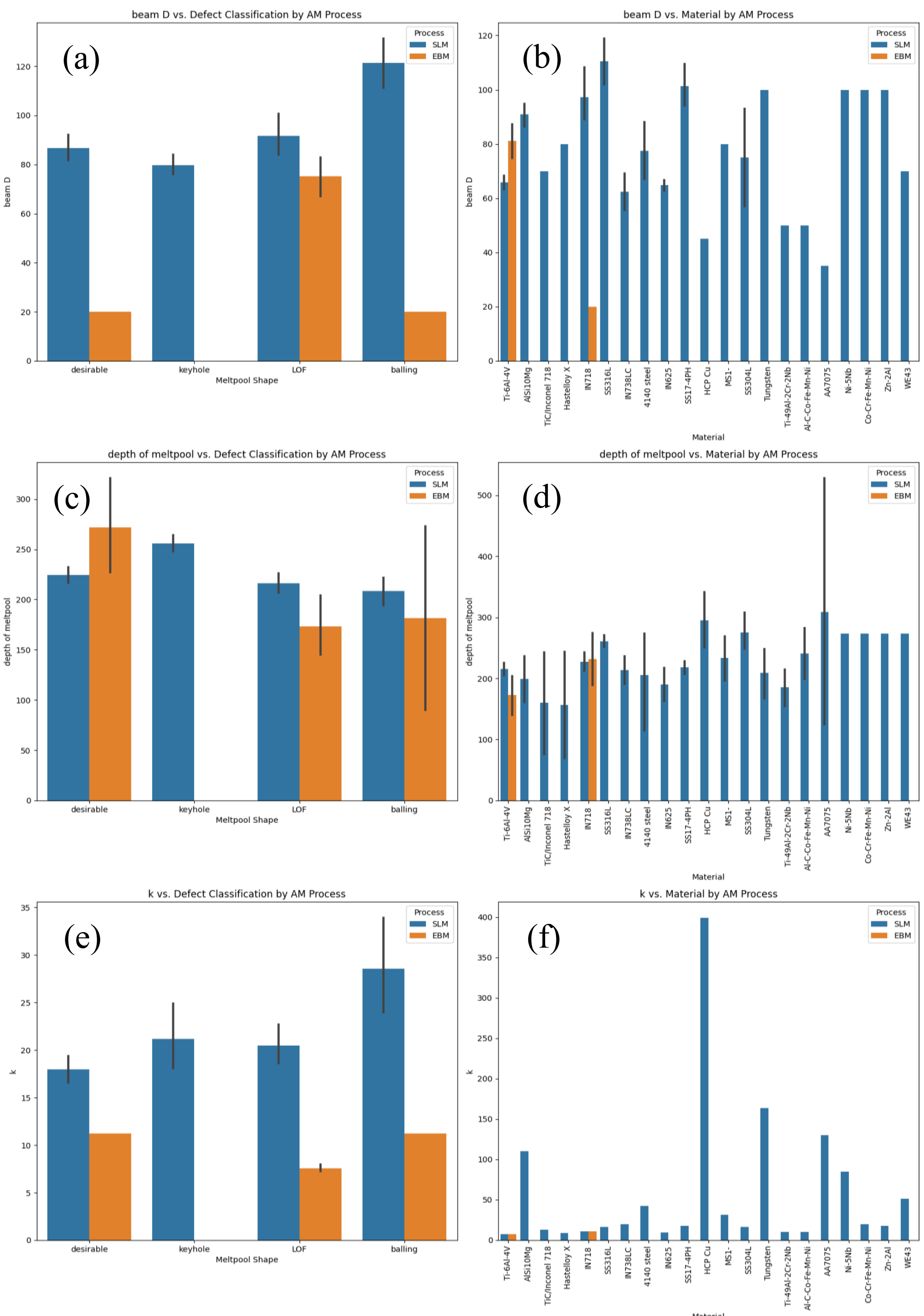

**Figure 5.** Distribution of Defect and Material classifications by sub-categorical AM processes as SLM and EBM in our datasets under **(a,b)** Laser Spot Diameter, **(c,d)** Depth of Melt Pool, and **(e,f)** Thermal Conductivity.

### *2.3 Featurization*

During the featurization process, we meticulously select and construct features from our dataset, comprising 1514 training datasets and 505 testing datasets, to feed into our ML models for prediction purposes. Given the intricate nature of AM and the multitude of geometrical and material properties inherent in melt pool phenomena, it's crucial to delineate a substantial number of features to train an effective ML model for property prediction. Considering that AM processes entail a blend of numerical and categorical features, incorporating both one hot encoding and sub-categorical features of laser/beam-based AM processes like Selective Laser Melting (SLM) and Electron Beam Melting (EBM) becomes imperative. One hot encoding facilitates the conversion of categorical features, where each value belongs to one of several non-numeric categories, into numeric categories comprehensible by conventional ML models. This encoding transforms a categorical feature with $n$ possibilities into $n$ binary encoding features. In this scheme, for a data sample belonging to a specific category, a 1 is assigned to the corresponding encoding feature, while 0 is assigned to the remaining $n-1$ encoding features. This approach enables us to undertake prediction tasks while recognizing the varying relationships that different heat sources and feedstock material supply methods may have with other features and the prediction target.

### *2.4 Dataset splitting*

For each dataset, we have designated a metric and a splitting pattern that aligns well with the dataset's properties. Given the highly heterogeneous nature of our dataset's input parameters, we perform scaling for both categorical factors such as material and process types, as well as sub-categorical processes, along with numerical inputs. Additionally, our output classes - balling, desirable, LOF, and keyhole - undergo normalization. The equation utilized for normal scaling in our benchmark is as follows:

$$x' = \frac{x - \min(x)}{\max(x) - \min(x)}$$

where, $x'$ is the scaled value, $x$ is the original value, $\min(x)$ is the minimum value in the dataset, and $\max(x)$ is the maximum value in the dataset.

In machine learning, datasets need to be divided into training and test subsets to assess the performance of models on unseen data. Accordingly, we partitioned our dataset into training and testing sets. The models were trained using the training data, while the test data was kept separate for evaluating model performance. This process was carried out after scaling the datasets, as illustrated in ***Figure 6***.

Furthermore, k-fold cross-validation is employed to gain deeper insights into model performance by considering multiple training and test partitions. In this method, the dataset is divided into k partitions. Subsequently, k−1 partitions are utilized for training the model, while the remaining partition is held out for testing. This process is repeated k times, ensuring that each partition is used for testing exactly once.

The primary aim of cross-validation is to mitigate overfitting, where a model is excessively tuned to the training data, resulting in poor performance on new, unseen data. By evaluating the model across various validation sets, cross-validation provides a more accurate estimation of the model's ability to generalize to new data. However, this approach can introduce increased variability in testing models since it evaluates against single data points. Outliers within these data points can significantly impact the variability of the testing process. Additionally, k-fold cross-validation can be computationally expensive due to its iteration over the number of data points. To address these concerns and prevent overfitting, we selected k=11 as it demonstrated highly effective performance with minimal errors and overfitting, as depicted in ***Figure 7***.

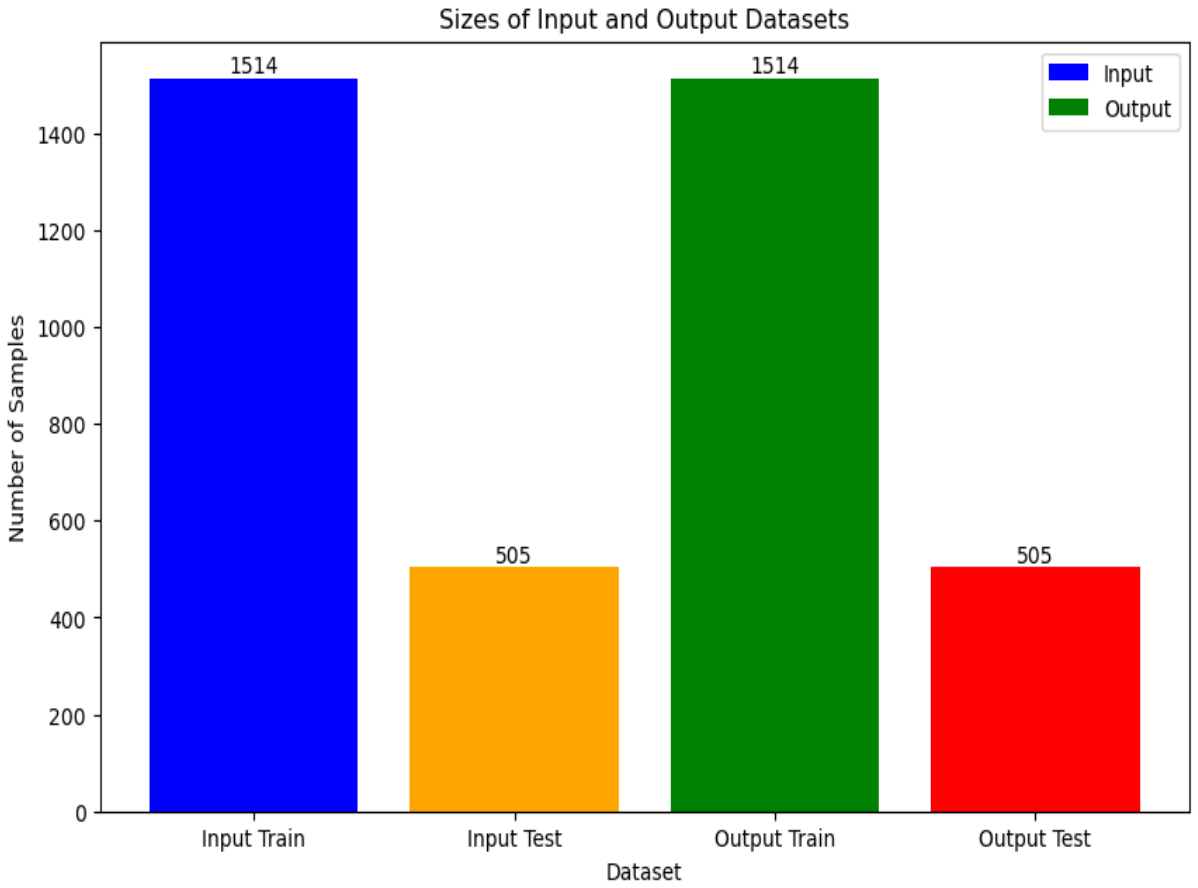

**Figure 6.** Split AM datasets as training and test for both input and output.

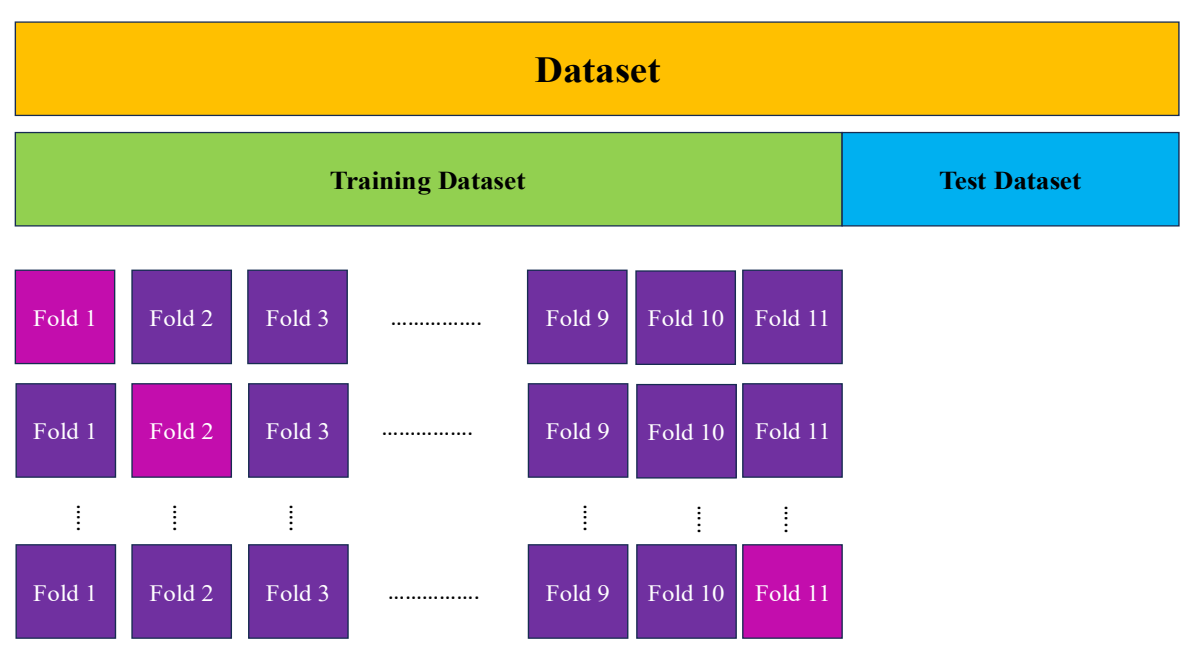

**Figure 7**. $k$-fold cross validation with $k$ =11 was employed on the AM datasets as a preparatory step before applying machine learning models.

## *2.5 ML Models*

Within the domain of multi-physics AM, time and cost considerations are pivotal, prompting the exploration of ML models as viable alternatives to traditional experimental and numerical methods. This study investigates the performance of various ML models using pre-segmented datasets as two main division of linear and non-linear algorithms. Below, we offer a succinct overview of these models, with a more elaborate discussion on effective metric methodologies in the ensuing section. Subsequently, the results and discussions section will unveil the outcomes of these models, followed by a comparative analysis in the metrics section to discern their respective efficiencies.

### 2.5.1 Linear Classification Algorithm

A classification algorithm, known as a classifier, determines its classifications based on a linear predictor function that combines a set of weights with the feature vector, as shown in the equation below:

$$y = f(\vec{w}.\vec{x}) = f(\sum_{j} w_j . x_j)$$

Here, $y$ represents the output, $f$ is the activation function, $\vec{w}$ denotes the weights, $\vec{x}$ is the feature vector, and $\vec{w}.\vec{x}$ represents the dot product of weights and features.

Given the widespread use of linear ML algorithms, this study focuses on employing two well-known models: logistic regression and support vector machine (SVM) with a linear kernel, also known as Support Vector Classification (SVC).

#### *2.5.1.1 Logistic Regression*

Logistic regression is a statistical method used to model the probability of a discrete outcome given an input variable. While commonly applied to binary classification tasks, it can also be extended to handle multiclass scenarios. In multiclass logistic regression, where there are more than two possible outcomes, the model utilizes multiple logistic functions, each corresponding to a specific class. The equation for multiclass logistic regression can be expressed as follows:

$$P(y_i = k|x_i) = \frac{e^{w_k.x_i}}{\sum_{j=1}^{k} e^{w_j.x_i}}$$

Here, $P(y_i = k|x_i)$ represents the probability that the input sample $x_i$, belongs to class $k$, $w_k$ is the weight vector corresponding to class $k$, and $x_i$ is the feature vector of the $i$-th sample, $k$ denotes the total number of classes.

#### *2.5.1.2 Support Vector Machine (SVM)*

##### 2.5.1.2.1 Linear Function Kernel

Support Vector Machines (SVMs) are a type of machine learning algorithm that predicts by identifying a hyperplane separating data points of different categories [14]. This hyperplane is positioned to maximize the margin, or distance, between the nearest data points and the decision boundary. In the case of classification with a linear kernel, the decision function of SVM can be expressed mathematically as follows:

$$f(x) = w.x + b$$

where, $f(x)$ represents the decision function that predicts the class label of an input sample $x$, $w$ is the weight vector that determines the orientation of the hyperplane, $x$ is the input feature vector and $b$ is the bias term.

### 2.5.2 Non-Linear Classification Algorithm

However, in scenarios where classes cannot be separated by a linear boundary, non-linear classifiers become essential in using machine learning models. These classifiers are adept at handling intricate classification challenges by capturing complex patterns and relationships within data. Unlike linear models, non-linear classifiers offer enhanced performance when faced with complex datasets. Below, we present the most common models fitted to our four classes of melt pool shape, encompassing tree-based algorithms, Neural Networks (NNs) and Gaussian Naive Bayes (GaussianNB) algorithms, providing insight into the capabilities of non-linear algorithms.

#### *2.5.2.1 Tree-based model*

Tree-based models, such as decision trees, random forests, and gradient boosting machines, are widely used in machine learning for both classification and regression tasks. These models operate by recursively partitioning the feature space into smaller regions based on feature values, with each split optimizing a chosen criterion, such as Gini impurity or information gain. This process results in a tree-like structure where each leaf node represents a final decision or prediction. Random forests and gradient boosting machines further enhance predictive performance by combining multiple trees to form robust ensemble models. Overall, tree-based models are valued for their interpretability, flexibility, and ability to capture complex relationships in the data.

#### 2.5.2.1.1 Decision Trees

Decision Tree is a non-parametric supervised learning method utilized for classification and regression tasks. In our context, we focus on its application for classifying four classes of melt pool shape. Essentially, it operates by iteratively partitioning the input space into regions based on the feature values to predict the target variable. At each internal node of the tree, a decision is made using a specific feature value, leading to multiple branches. This process continues until a leaf node is reached, where the prediction for the target variable is made.

In the case of multiple classes, the Decision Tree extends its binary concept to handle scenarios with more than two outcomes. Instead of dividing the dataset into two branches at each node, a multi-class decision tree divides it into multiple branches, each corresponding to one possible class. The decision-making process involves recursively partitioning the feature space based on feature values, with each split aiming to maximize the purity of resulting subsets in terms of class labels. This concept is illustrated below:

$$\begin{cases} class_1 & if\ Feature < threshold_1 \\ class_2 & if\ threshold_1 \leq Feature < threshold_2 \\ class_n & if\ threshold_{n-1} \leq Feature < threshold_n \end{cases}$$

#### 2.5.2.1.2 Random Forest

Random Forest is a powerful tree-based model used for both classification and regression tasks. It operates by constructing a multitude of decision trees during training and outputs the mode of the classes (classification). Each tree in the forest is trained on a random subset of the training data and a random subset of the features. This randomness helps to decorrelate the trees, making the model less prone to overfitting and more robust. In our classification tasks, Random Forest aggregates the predictions of individual trees to determine the final class label. It is particularly effective in handling high-dimensional data and is less sensitive to noisy features and outliers compared to a single decision tree. The algorithm's ability to capture complex relationships in the data and its resilience to overfitting make it a popular choice for various machine learning tasks.

#### 2.5.2.1.3 Gradient Boosting Trees (GBT)

Gradient Boosting Trees (GBT) is a robust tree-based algorithm that optimizes a loss function by iteratively adding decision trees to the ensemble. In each iteration, a new tree is trained to predict the residuals, which are the differences between the actual and predicted values, of the previous trees. These predictions are then combined to generate the final prediction. GBT excels in handling complex relationships within the data and is renowned for its capability to produce highly accurate predictions. It particularly shines in scenarios where other machine learning algorithms face challenges, such as when dealing with noisy or high-dimensional data.

##### 2.5.2.1.3.1 Extreme Gradient Boosting Machine (XGBM)

Extreme Gradient Boosting Machine (XGBoost) is an advanced implementation of the gradient boosting algorithm that has gained popularity for its speed and performance. It is designed to optimize the gradient boosting process, incorporating features such as parallel computing, regularization, and tree pruning to enhance accuracy and efficiency.

##### 2.5.2.1.3.2 Light Gradient Boosting Machine (LGBM)

The Light Gradient Boosting Machine (LGBM) is a gradient boosting framework similar to XGBoost, renowned for its speed, efficiency, and scalability. It excels in handling large-scale datasets and has gained widespread popularity across various machine learning tasks due to its outstanding performance. LGBM is celebrated for its rapid training speed and efficiency, attributes attributed to its leaf-wise tree growth strategy and optimizations like Gradient-Based One-Side Sampling (GOSS) and Exclusive Feature Bundling (EFB). In contrast, XGBoost employs a depth-wise tree growth strategy by default, which may result in slower performance, particularly with sizable datasets.

##### 2.5.2.1.3.3 Adaptive Gradient Boosting Machine (AdaBoost)

AdaBoost, short for Adaptive Boosting, is an ensemble learning method that builds a strong classifier by combining multiple weak classifiers. It works by sequentially training a series of weak learners on weighted versions of the training data. In each iteration, the algorithm focuses more on the instances that were misclassified in the previous iteration, effectively adjusting its approach to improve performance.

AdaBoost assigns a weight $\alpha_t$ to each weak learner $h_t(x)$, where $t$ represents the iteration number. The final prediction $H(x)$ is then obtained by summing the weighted predictions of all weak learners:

$$H(x) = sign(\sum_{t=1}^{T} \alpha_t h_t(x))$$

Here, $T$ denotes the total number of weak learners. The sign function ensures that the final prediction is either +1 or -1, depending on the overall weighted sum of the weak learners' predictions.

###### 2.5.2.1.3.4 Categorical Gradient Boosting Machine (CatBoost)

Categorical Gradient Boosting Machine (CatBoost) is a gradient boosting algorithm designed to handle categorical features seamlessly which is similar to other gradient boosting algorithms like XGBoost and LightGBM but incorporates specific optimizations to effectively deal with categorical data without requiring pre-processing steps like one-hot encoding. CatBoost introduces a novel approach called ordered boosting, which optimizes the sequence of trees added to the ensemble, leading to improved performance. The algorithm works by iteratively training decision trees on the dataset, where each tree is built to minimize a specified loss function.

###### 2.5.2.1.3.5 Bagging (Bootstrap Aggregating)

Bagging, short for Bootstrap Aggregating, is an ensemble learning method that aims to improve the stability and accuracy of machine learning models by reducing variance and overfitting. It works by training multiple instances of the same base learning algorithm on different subsets of the training data and then combining their predictions through a process called aggregation, the aggregation process can be represented as follows:

$$\hat{f}_{bag}(x) = \frac{1}{B} \sum_{b=1}^{B} \hat{f}_b(x)$$

where, $\hat{f}_{bag}(x)$ represents the aggregated prediction for the instance $x$, obtained by averaging the predictions $\hat{f}_b(x)$ from each individual model $b$, where $B$ is the total number of models.

###### 2.5.2.1.3.6 Voting

In our study, we utilize voting as a technique to amalgamate predictions from multiple individual models, specifically Random Forest and Gradient Boosting. By combining these two methods, each of which demonstrates strong classification capabilities independently, we aim to achieve a comprehensive assessment of their collective performance, the aggregation process in soft voting can be represented as:

$$\hat{f}_{Voting}(x) = \frac{1}{M} \sum_{m=1}^{M} \hat{f}_m(x)$$

Where, $\hat{f}_{Voting}(x)$ represents the final prediction for the instance x, obtained by averaging the predicted $\hat{f}_m(x)$ probabilities from each individual model $m$, and $M$ is the total number of models.

#### *2.5.2.2 Neural Networks (NNs)*

Neural Networks (NNs) are versatile machine learning algorithms suitable for both regression and classification tasks [15]. Within these networks, individual neurons perform linear and nonlinear transformations on input data, producing outputs that are adjusted through the iterative process of backpropagation, wherein weights and biases are updated to optimize model performance.

#### *2.5.3.1 Multilayer Perceptrons (MLPs)*

Multilayer Perceptrons (MLPs) represent a class of neural networks distinguished by their layered architecture comprising interconnected neurons. These networks typically consist of an input layer, one or more hidden layers, and an output layer. Within the network, each neuron processes input data through weighted connections and applies activation functions to generate output. MLPs find extensive applications in diverse machine learning tasks such as classification, regression, and pattern recognition due to their capability to capture intricate data relationships. The general equation governing the behavior of a node in an MLP is as follows:

$$a_i = f\left(\sum_{j=1}^{n} w_{ij} . x_j + b_i\right)$$

where, $a_i$ is the output of the $i$-th node in the layer, $f$ is the activation function applied element-wise, $w_{ij}$ is the weight connecting the $j$-th input to the $i$-th node, $x_j$ is the $j$-th input to the node, $b_i$ is the $i$-th bias term for the node, and $n$ is the number of inputs to the node.

#### *2.5.2.3 Support Vector Machine (SVM)*

###### 2.5.2.3.1 Radial Basis Function (RBF) Kernel

Support Vector Classification (SVC) with Radial Basis Function (RBF) Kernel is a variant of Support Vector Machines (SVM), which is a powerful supervised learning algorithm used for classification tasks [16]. The RBF kernel is particularly effective in handling non-linear relationships between features in the dataset. It works by transforming the input space into a higher-dimensional space where the classes can be more easily separated by a hyperplane. This transformation is achieved using a Gaussian radial basis function. The RBF kernel has two main parameters: gamma (γ) and regularization parameter (C), which control the flexibility of the decision boundary and the trade-off between maximizing the margin and minimizing classification errors, respectively.

#### 2.5.2.3.2 Polynomial Function Kernel

In SVC, Polynomial Function Kernel as another type of SVM, data points are mapped to a higher-dimensional space using polynomial transformations, allowing for the creation of non-linear decision boundaries. The polynomial kernel function computes the dot product between pairs of data points in the transformed space, enabling effective classification in cases where the relationship between features and classes is non-linear. The polynomial kernel function is defined as:

$$K(x, y) = (x^T y + c)^d$$

where $x, y$ are input feature vectors, $c$ is a constant term, and $d$ is the degree of the polynomial. This kernel function allows SVC to capture complex patterns and achieve high accuracy in classification tasks.

#### 2.5.2.3.3 Sigmoid Function Kernel

Support Vector Classifier (SVC) with Sigmoid Function Kernel is another variant of the support vector machine (SVM) algorithm, which utilizes a sigmoid function as its kernel. The sigmoid kernel function is defined as:

$$K(x, y) = \tanh(\alpha x^T y + c)$$

where, $x, y$ are input feature vectors, $\alpha$ and $c$ are constants, and $\tanh$ is the hyperbolic tangent function. The sigmoid kernel function allows SVC to create non-linear decision boundaries by transforming the input space into a higher-dimensional space. It is particularly useful for classification tasks where the relationship between features and classes is non-linear. However, it may be more sensitive to noise and less robust compared to other kernel functions like the polynomial or radial basis function (RBF) kernels.

### *2.5.2.4 Instance-based learning algorithm (Lazy Learner)*

Instance-based learning, also known as lazy learning, is a type of machine learning algorithm where the system learns by comparing new instances with stored instances, rather than through explicit generalization. In lazy learning, the algorithm does not build a model during the training phase. Instead, it stores the entire training dataset and waits until a new instance needs to be classified or predicted. When a prediction is required, the algorithm retrieves the most similar instances from the training data and uses them to make a prediction for the new instance. One of the most popular instance-based learning algorithms is the k-Nearest Neighbors (k-NN) algorithm. In k-NN, the algorithm classifies a new instance based on the class labels of its k nearest neighbors in the training data. Other instance-based algorithms include locally weighted learning (LWL) and Case-Based Reasoning (CBR), however, in this study we used k-NN model.

#### 2.5.2.4.1 k-Nearest Neighbors (k-NN) algorithm

The k-Nearest Neighbors (k-NN) algorithm is a type of instance-based learning method used for both classification and regression tasks in machine learning. In k-NN, the prediction for a new data point is determined by the majority class (for classification) or the average value (for regression) of its k nearest neighbors in the feature space.

## *2.6 Evaluation Metrics*

To assess the effectiveness of our machine learning models on new data, we initially randomized our datasets and then conducted 11-fold cross-validation. This technique involves dividing the data into 11 subsets and, over 11 rounds, using one subset for validation while the others serve as the training set. The overall accuracy of our models is determined by averaging the accuracies obtained across all eleven iterations.

Our dataset focuses on a classification task centered on identifying melt pool defects in AM, termed AM-built defect detection. Various evaluation metrics are employed, including accuracy, precision (micro-computes across all classes, macro-computes for each class and averages them, treating all classes equally, and weighted- Computes the average precision weighted by the number of instances in each class), recall (micro, macro, and weighted), and F1 score (micro, macro, and weighted). These metrics offer comprehensive insights into the model's performance, considering factors such as class imbalance and overall effectiveness across all classes. To visualize, additionally, the learning progress of our models, we employ learning curves. These curves illustrate how specific metrics, such as accuracy or loss, evolve throughout the model training process, providing insights into the model's performance over successive iterations.

## 3. Results and discussion

In this section, we analyze the performance of AM-DefectNet benchmarked models on datasets. Initially, the datasets were collected, cleaned, and prepared by removing illogical data and applying methods like forward, backward, and polynomial filling. Subsequently, each model's efficacy was assessed, comparing linear and non-linear architectures. Our results are presented using various metrics and learning curves, with hyperparameter optimization to enhance model accuracy. This work represents a significant advancement in ML application for AM investigation, providing extensive comparisons previously unseen in the field. This investigative methodology will contribute to a deeper understanding of ML's role in classifying AM-built defects. Further details will be discussed in subsequent sections.

### *3.1 Classification Task*

In the AM-DefectNet classification task, we explored the effectiveness of 15 machine learning algorithms. These algorithms include Random Forest, Support Vector Classifier (with linear, sigmoid, rbf, and polynomial kernels), Logistic Regression, and Gradient Boosting techniques such as XGBoost, AdaBoost, CatBoost, and LGBM. Additionally, we employed Neural Networks, specifically Multi-Layer Perceptrons (MLPs).

Our investigation focused on predicting melt pool Laser Powder Bed Fusion (LPBF) classifications across four classes: Keyhole, Desirable, Balling, and Lack of Fusion (LOF). We evaluated the performance of these models using various metrics including accuracy, precision, recall, F1-score, and confusion matrices. These evaluations were conducted on both test and unseen datasets, categorized into macro, micro, and weighted classifications. In our analysis, a macro-average was utilized to independently compute metrics for each class, followed by averaging the results. On the other hand, a micro-average aggregated the contributions of all classes to compute a unified average metric. Additionally, the weighted approach was employed, where weight values, ranging between zero and one, were assigned to each class, ensuring a normalized rating with a total value of one. The comparison of all ML models used in this study is presented in ***Figure 8*** and ***Table 1***, showcasing their performance across different evaluation metrics and dataset categories. Additionally, we utilized confusion matrices to visually depict how well our implemented models performed. Each matrix represents the distribution of model predictions across different classes, compared to the actual occurrences in the ground truth data. Through label encoding, where 'LOF' corresponds to 0, 'balling' to 1, 'desirable' to 2, and 'keyhole' to 3, we evaluated the performance of all 15 machine learning models on unseen datasets.

Within the context of gradient boost algorithms, as illustrated in Figure 11 (a-d), CatBoost algorithm emerged as the most effective when compared to XGBoost, LGBM, and AdaBoost, respectively. This superiority is evident in various performance metrics such as precision, recall, F1-score, and support. The CatBoost algorithm exhibited strong performance across various metrics. In terms of precision, which measures the accuracy of positive predictions, CatBoost achieved values ranging from 0.89 to 0.97 across different classes. This indicates that the algorithm accurately identified instances of each class. Furthermore, the recall metric, which represents the ability of the model to correctly identify true positives, ranged from 0.88 to 0.96 for CatBoost. This suggests that the algorithm effectively captured the majority of instances belonging to each class. The F1-score, which is the harmonic mean of precision and recall, ranged between 0.90 and 0.95 for CatBoost. This metric provides a balanced assessment of both precision and recall, indicating the overall effectiveness of the algorithm in classification tasks. Additionally, the support metric denotes the number of occurrences of each class in the dataset. CatBoost demonstrated robust support across all classes, with values ranging from 54 to 190. Overall, CatBoost achieved an accuracy of 0.92, indicating the proportion of correctly classified instances out of the total dataset. This highlights the algorithm's ability to accurately classify instances across all classes. The complexity of the AM process posed challenges for linear classification algorithms, while tree-based models with non-linear capabilities exhibited superior performance. As illustrated in ***Figure 10***, logistic regression struggled with classifying AM defects, resulting in higher errors on unseen datasets. In contrast, Decision Tree, Random Forest, and K-NN models showcased better performance, as depicted in ***Figure 10*** (b-d).

.

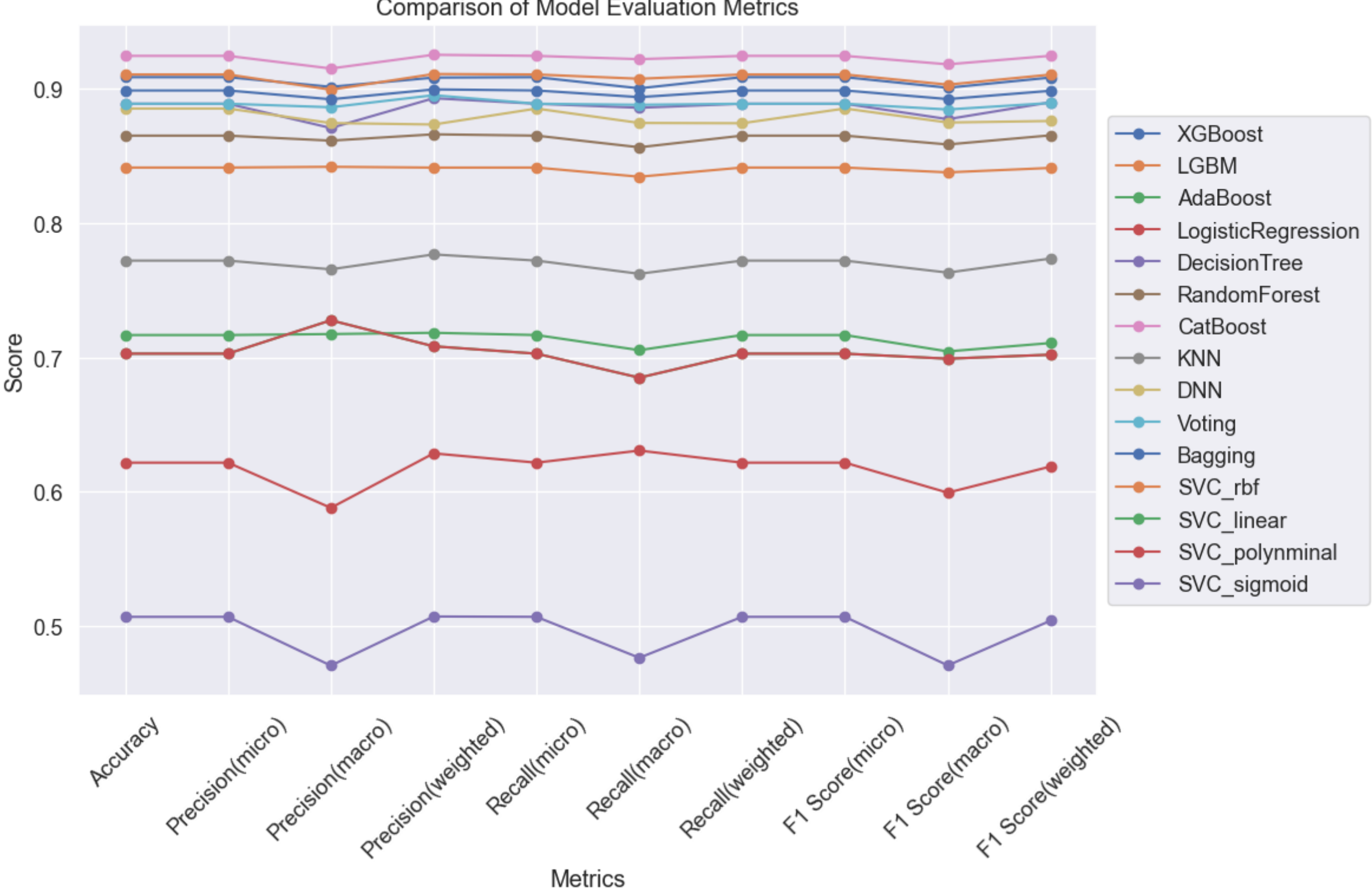

**Figure 8.** Comparison of ML Model's Evaluation.

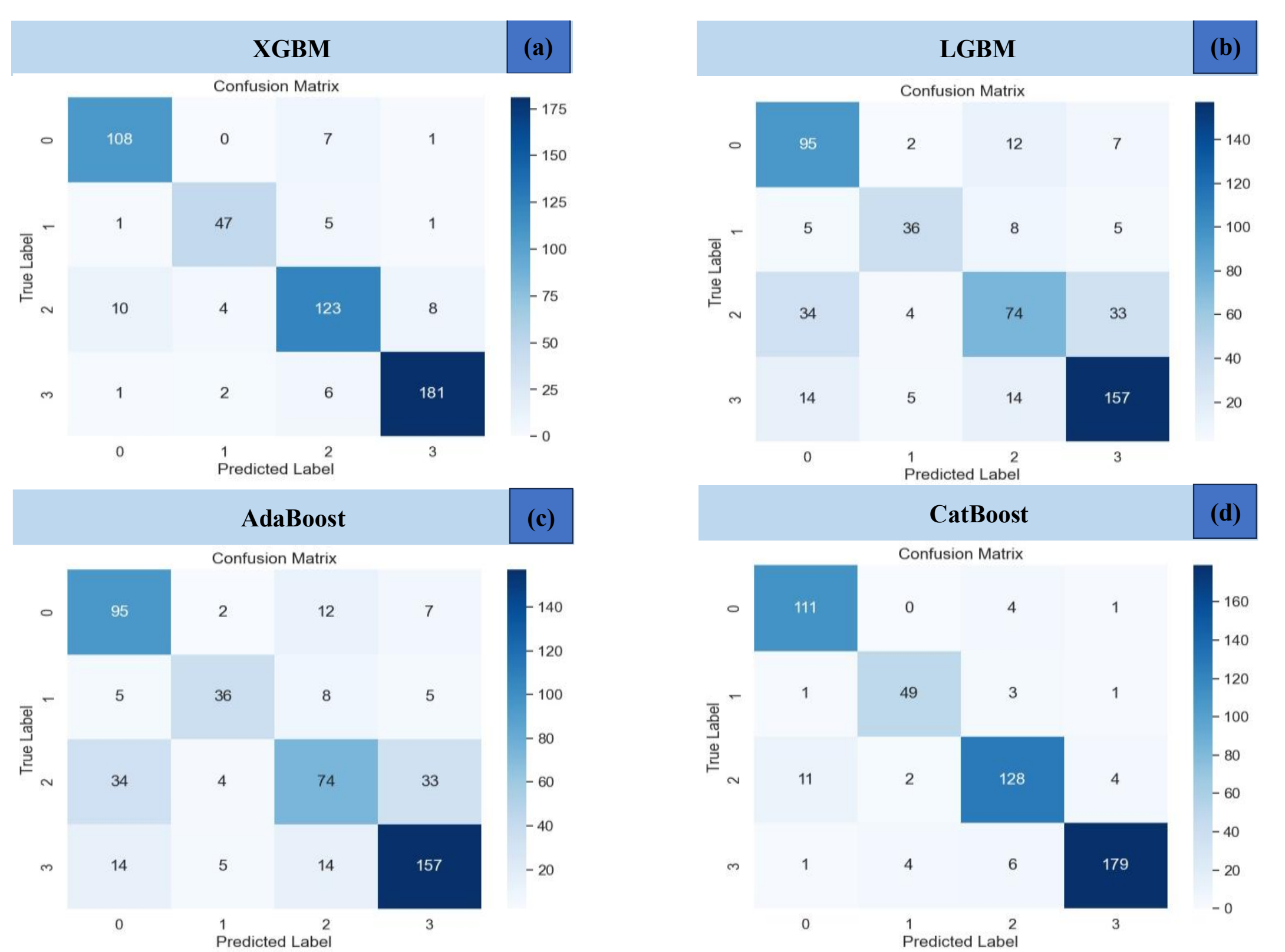

**Figure 9.** Confusion Matrix. (a) XGBM, (b)LGBM, (c)AdaBoost, and (d) CatBoost Models.

**Table 1.** The evaluation of ML models involved the assessment of four key metrics: accuracy, precision, recall, and F1-score.

| ML Models | | | Metrics | |
|---|---|---|---|---|
| **Linear Classification Algorithm** | Logistic Regression | | Accuracy | 0.6217 |
| | | | Precision | 0.6217 |
| | | | Recall | 0.6217 |
| | | | F1 Score | 0.6217 |
| | SVR+Linear Kernel | | Accuracy | 0.7029 |
| | | | Precision | 0.7029 |
| | | | Recall | 0.7029 |
| | | | F1 Score | 0.7029 |
| **Non-Linear Classification Algorithm** | Tree-based model | Decision Trees | Accuracy | 0.8891 |
| | | | Precision | 0.8891 |
| | | | Recall | 0.8891 |
| | | | F1 Score | 0.8891 |
| | | Random Forest | Accuracy | 0.8653 |
| | | | Precision | 0.8653 |
| | | | Recall | 0.8653 |
| | | | F1 Score | 0.8653 |
| | | XGBM | Accuracy | 0.9089 |
| | | | Precision | 0.9089 |
| | | | Recall | 0.9089 |
| | | | F1 Score | 0.9089 |
| | | LGBM | Accuracy | 0.9108 |
| | | | Precision | 0.9108 |
| | | | Recall | 0.9108 |
| | | | F1 Score | 0.9108 |
| | | AdaBoost | Accuracy | 0.7168 |
| | | | Precision | 0.7168 |
| | | | Recall | 0.7168 |
| | | | F1 Score | 0.7168 |
| | | CatBoost | Accuracy | 0.9247 |
| | | | Precision | 0.9247 |
| | | | Recall | 0.9247 |
| | | | F1 Score | 0.9247 |
| | | Bagging | Accuracy | 0.8990 |
| | | | Precision | 0.8990 |
| | | | Recall | 0.8990 |
| | | | F1 Score | 0.8990 |
| | | Voting | Accuracy | 0.8891 |
| | | | Precision | 0.8891 |
| | | | Recall | 0.8891 |
| | | | F1 Score | 0.8891 |
| | Neural Networks (NNs) | MLPs | Accuracy | 0.8855 |
| | | | Precision | 0.8855 |
| | | | Recall | 0.8855 |
| | | | F1 Score | 0.8855 |
| | SVM | SVR+ RBF kernel | Accuracy | 0.8415 |
| | | | Precision | 0.8415 |
| | | | Recall | 0.8415 |
| | | | F1 Score | 0.8415 |
| | | SVR+ Polynomial kernel | Accuracy | 0.7029 |
| | | | Precision | 0.7029 |
| | | | Recall | 0.7029 |
| | | | F1 Score | 0.7029 |
| | | SVR+ Sigmoid kernel | Accuracy | 0.5069 |
| | | | Precision | 0.5069 |
| | | | Recall | 0.5069 |
| | | | F1 Score | 0.5069 |
| | lazy learner | k-NN | Accuracy | 0.7722 |
| | | | Precision | 0.7722 |
| | | | Recall | 0.7722 |
| | | | F1 Score | 0.7722 |

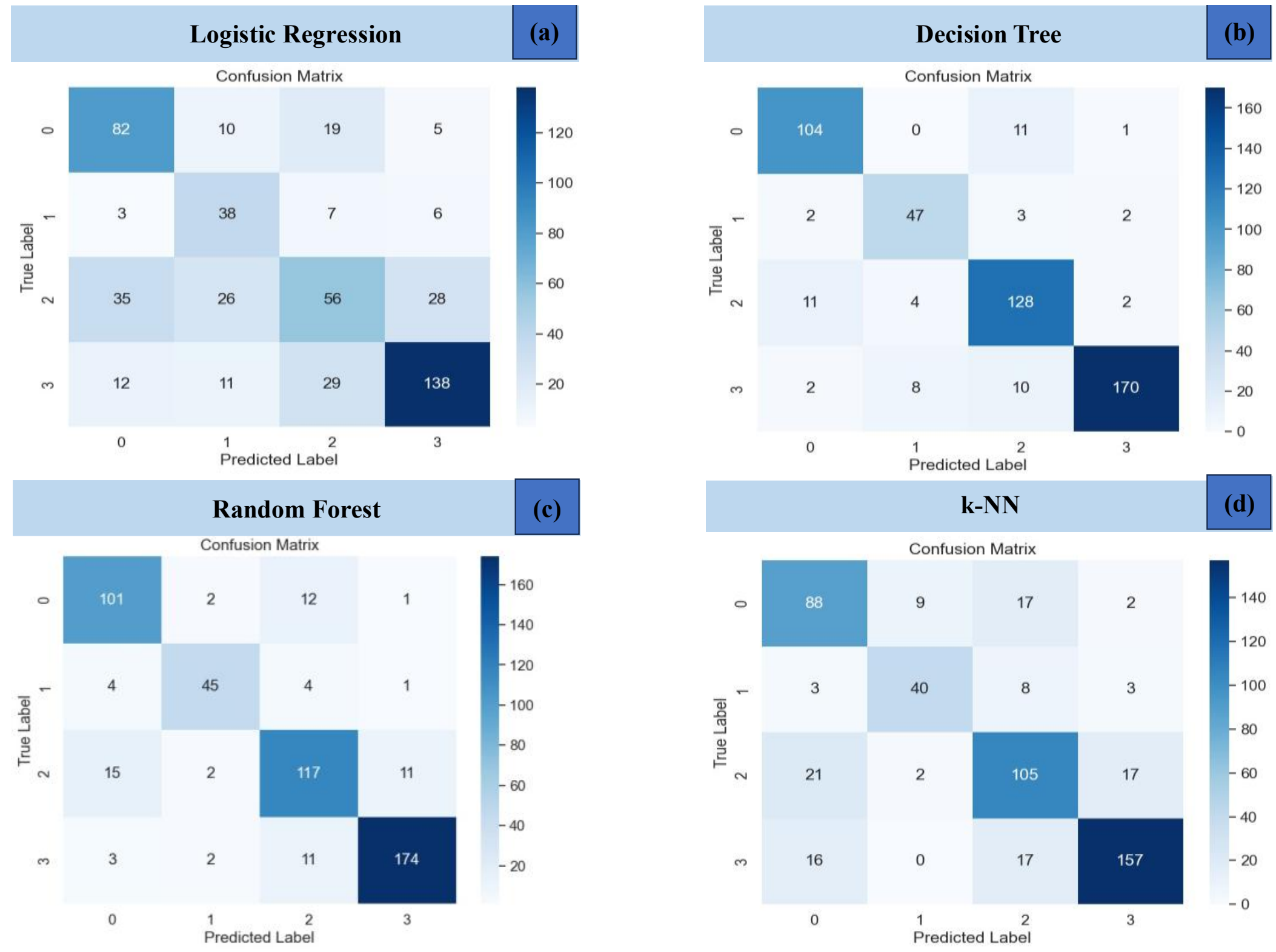

**Figure 10.** Confusion Matrix. (a) Logistic Regression, (b)Decision Tree, (c) Random Forest, and (d) k-NN Models.

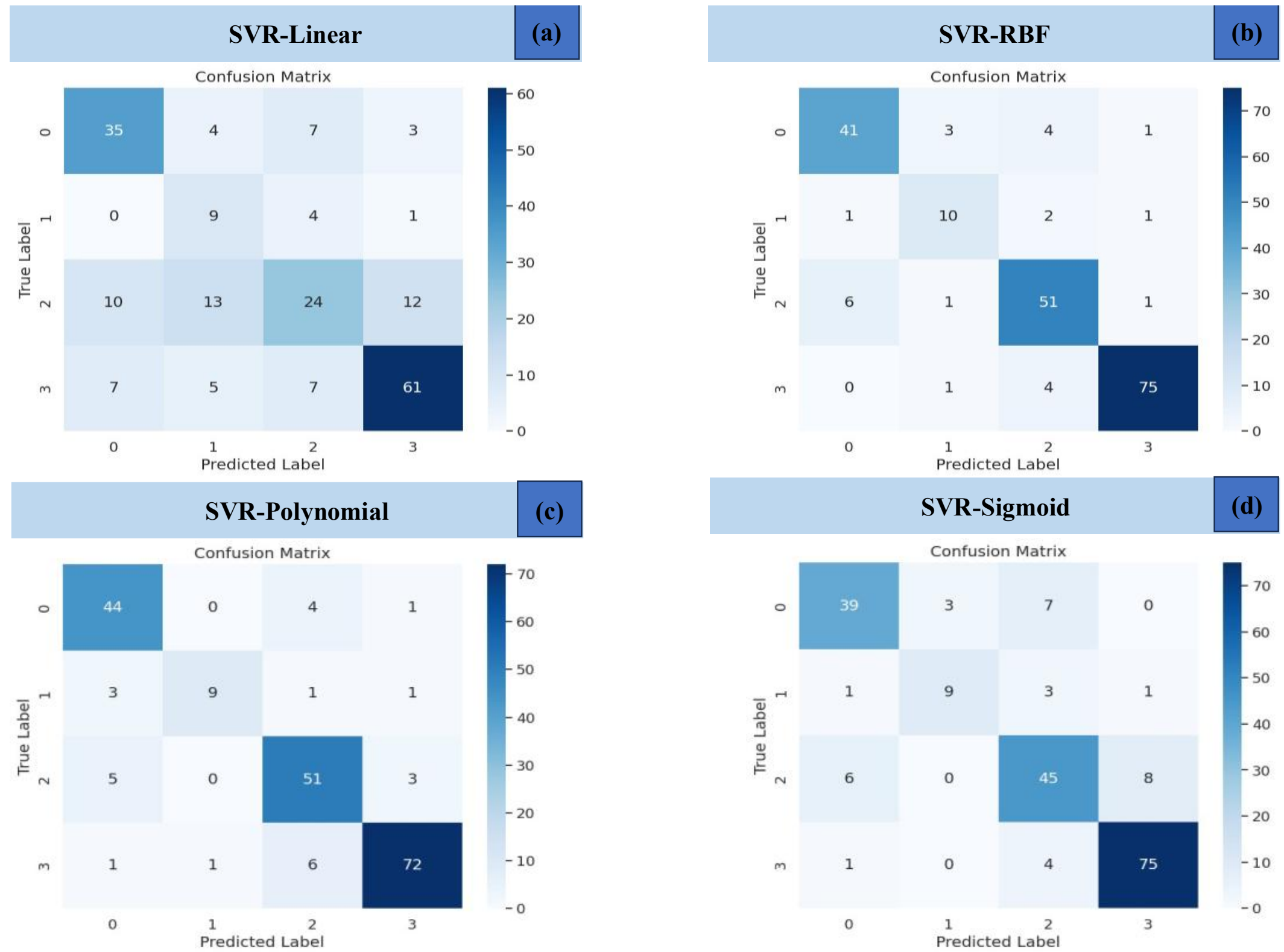

**Figure 11.** Confusion Matrix of SVM Models. (a) Linear, (b) RBF, (c) Polynomial, and (d) Sigmoid Kernels.

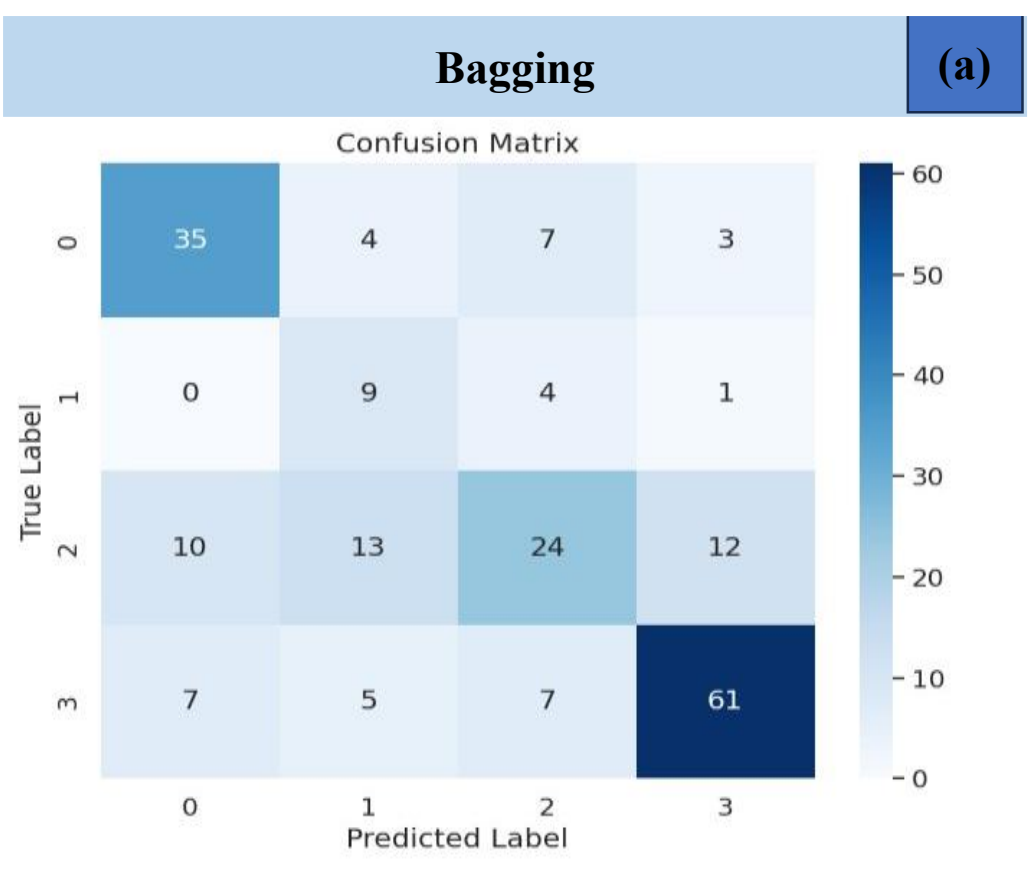

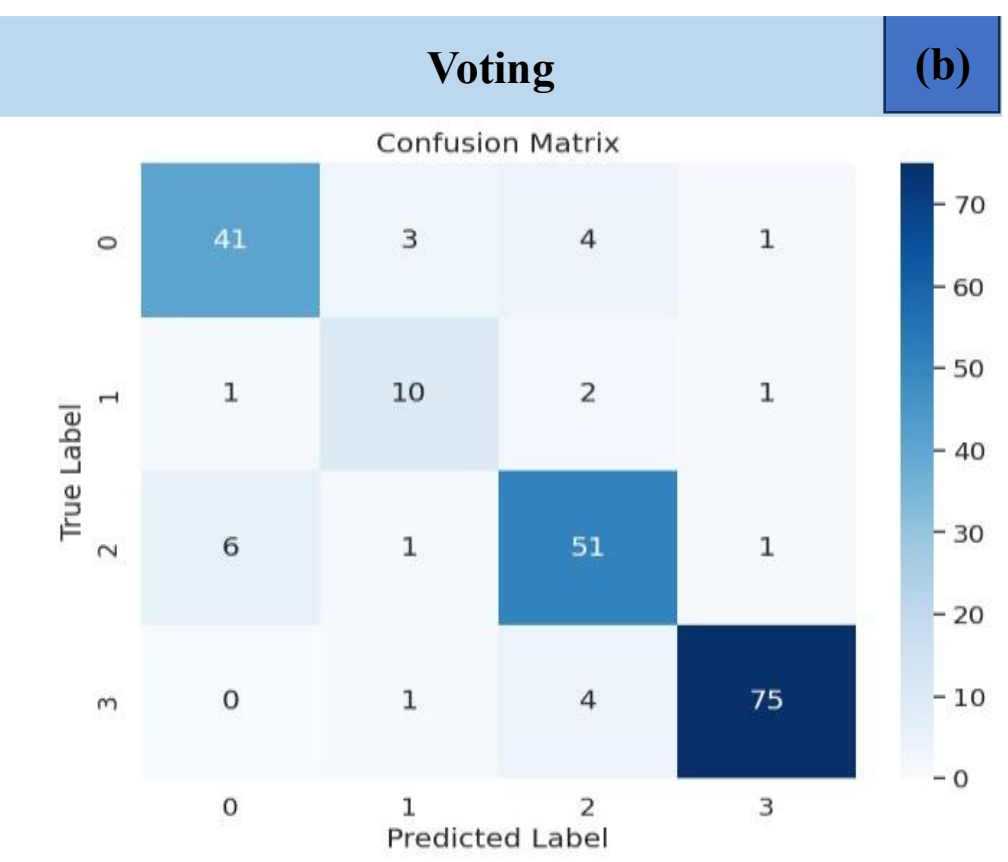

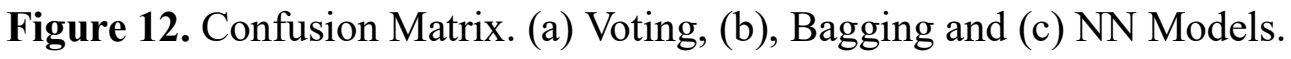

**Figure 12.** Confusion Matrix. (a) Voting, (b), Bagging and (c) NN Models.

Furthermore, among Support Vector Classifier (SVC) models illustrated in ***Figure 11***(a-d), the SVC with RBF kernel demonstrated superior performance compared to linear kernels, highlighting the ineffectiveness of linear classification algorithms for AM parameters.

Ensemble models such as Bagging and Voting, which combine random forest and gradient boost algorithms, revealed the superior performance of Voting in ***Figure 12*** (a, b). However, employing the Deep Neural Network (DNN) model for such complexity necessitated careful hyperparameter tuning. Our model, comprising 9 hidden layers, batch normalization, and l1/l2 regularization along with dropout to mitigate overfitting, did not yield satisfactory results. Nonetheless, Multi-Layer Perceptron (MLP) models, when compared to other published works, reported efficient results in training, validation, and test datasets, as shown in ***Figure 12*** (c). However, a logical overfitting is visible in more complex scenarios with insufficient samples, as discussed in the learning curve section.

### *3.2 Learning Curve*

A learning curve provides valuable insights into how the performance of a machine learning model evolves with the increasing amount of training data. It plots the training score and cross-validated test score against the number of training samples. Understanding the dynamics of learning curves can help in making informed decisions about model training and data requirements. Here's an elaboration on the points mentioned:

A. Convergence of Training and Cross-Validation Scores:

When the training and cross-validation scores converge as more data is added, it suggests that the model is reaching its optimal performance with the available data. In this scenario:

a) High Test Variability and Low Score: If there's high test variability and a consistently low score, it indicates that the model is not able to generalize well to unseen data, even with more training examples. This could be due to the complexity of the model or inherent noise in the data.
b) Low Test Variability and High Score: Conversely, if there's low test variability and a high score, it signifies that the model is performing well and consistently on the validation data. In such cases, adding more data may not significantly improve the model's performance.

B. Training Score Much Greater Than Validation Score:

When the training score significantly surpasses the validation score, it suggests that the model is overfitting to the training data. In other words, the model is capturing noise and patterns specific to the training set, which may not generalize well to new, unseen data. In such cases, adding more training examples can help the model generalize more effectively by providing it with more diverse instances to learn from.

### *3.3 Analysis and Validation*

In this section, we compare our results with prior research in the field of additive manufacturing and machine learning to validate the findings of our study and provide additional context for our conclusions. Among the top-performing models based on accuracy, the leading eight include CatBoost with an accuracy of 92.47%, followed closely by LGBM with 91.08%, and XGBoost with 90.89%. Following these tree-based models, Bagging achieved an accuracy of 89.9%, while both Voting and Decision Tree models demonstrated an accuracy of 88.91%. Additionally, Random Forest exhibited a commendable accuracy of 86.53%. It's noteworthy that the top-performing models primarily consist of non-linear tree-based algorithms. Alongside these, the Deep Neural Network (DNN) model, specifically the Multi-Layer Perceptron (MLP), displayed competitive performance with an accuracy of 88.51%, as shown in ***Figure 17***.

In order to gain deeper insights into the performance of our models, we have provided ***Table 3***, which includes the results from recent work. This comparison enables us to validate and assess the true performance of our models against existing benchmarks.

It is notable that while our models demonstrated improvement across a variety of algorithms, we observed no significant enhancement in the performance of SVM with four kernels. However, this improvement was evident in the performance of the other 11 algorithms across both numerical and categorical datasets used for AM defect classification. This discrepancy serves as a valuable indicator of the effectiveness of our AM-DefectNet benchmark in evaluating and enhancing model performance.

**Table 2.** Learning Curve details for ML Models.

| **ML Model** | **Model Performance** | **Figure Number** |
|---|---|---|
| Logistic Regression | High Test Variability and Low Score | ***Figure 14(a)*** |
| SVR+Linear Kernel | High Test Variability and Low Score | ***Figure 15(a)*** |
| Decision Trees | Additional data could enhance the model's already high performance, as it has not yet reached convergence between training and test scores. | ***Figure 14(b)*** |
| Random Forest | Additional data could enhance the model's already high performance, as it has not yet reached convergence between training and test scores. | ***Figure 14(c)*** |
| XGBM | Additional data could enhance the model's already high performance, as it has not yet reached convergence between training and test scores. | ***Figure 13(a)*** |
| LGBM | Low Test Variability and High Score | ***Figure 13(b)*** |
| AdaBoost | Low Test Variability and High Score | ***Figure 13(c)*** |
| CatBoost | Additional data could enhance the model's already high performance, as it has not yet reached convergence between training and test scores. | ***Figure 13(d)*** |
| Bagging | Low Test Variability and High Score | ***Figure 16(a)*** |
| Voting | Additional data could enhance the model's already high performance, as it has not yet reached convergence between training and test scores. | ***Figure 16(b)*** |
| MLPs | Good-fit model with an overfit started from 200 epoch | ***Figure 16(c)*** |
| SVR+ RBF kernel | Additional data could enhance the model's already high performance, as it has not yet reached convergence between training and test scores. | ***Figure 15(b)*** |
| SVR+ Polynomial kernel | Additional data could enhance the model's already high performance, as it has not yet reached convergence between training and test scores. | ***Figure 15(c)*** |
| SVR+ Sigmoid kernel | Additional data could enhance the model's already high performance, as it has not yet reached convergence between training and test scores. | ***Figure 15(d)*** |

| k-NN | Additional data could enhance the model's already high performance, as it has not yet reached convergence between training and test scores. | ***Figure 14(d)*** |
|---|---|---|

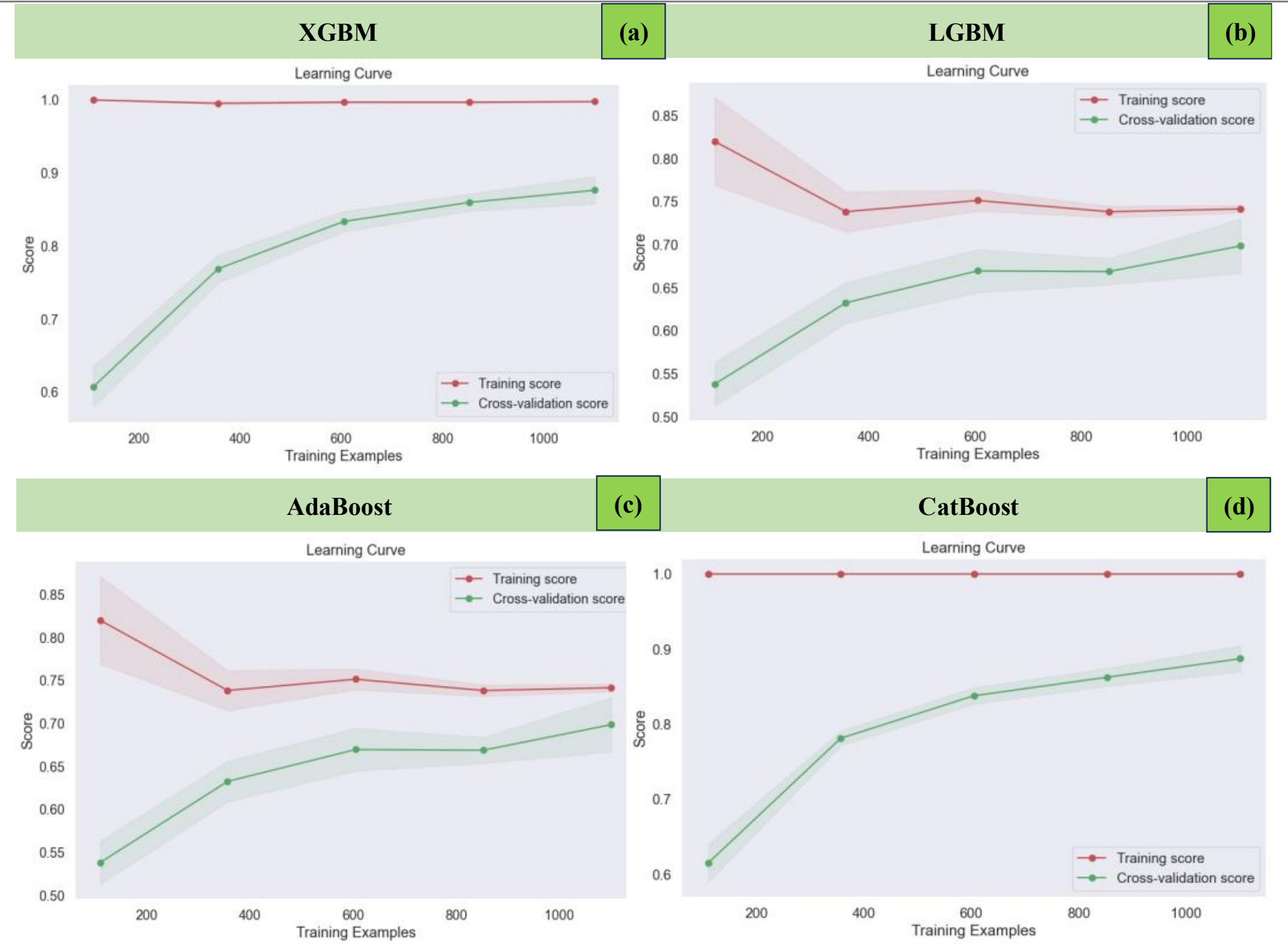

**Figure 13.** Learning Curve. (a) XGBM, (b)LGBM, (c)AdaBoost, and (d) CatBoost Models.

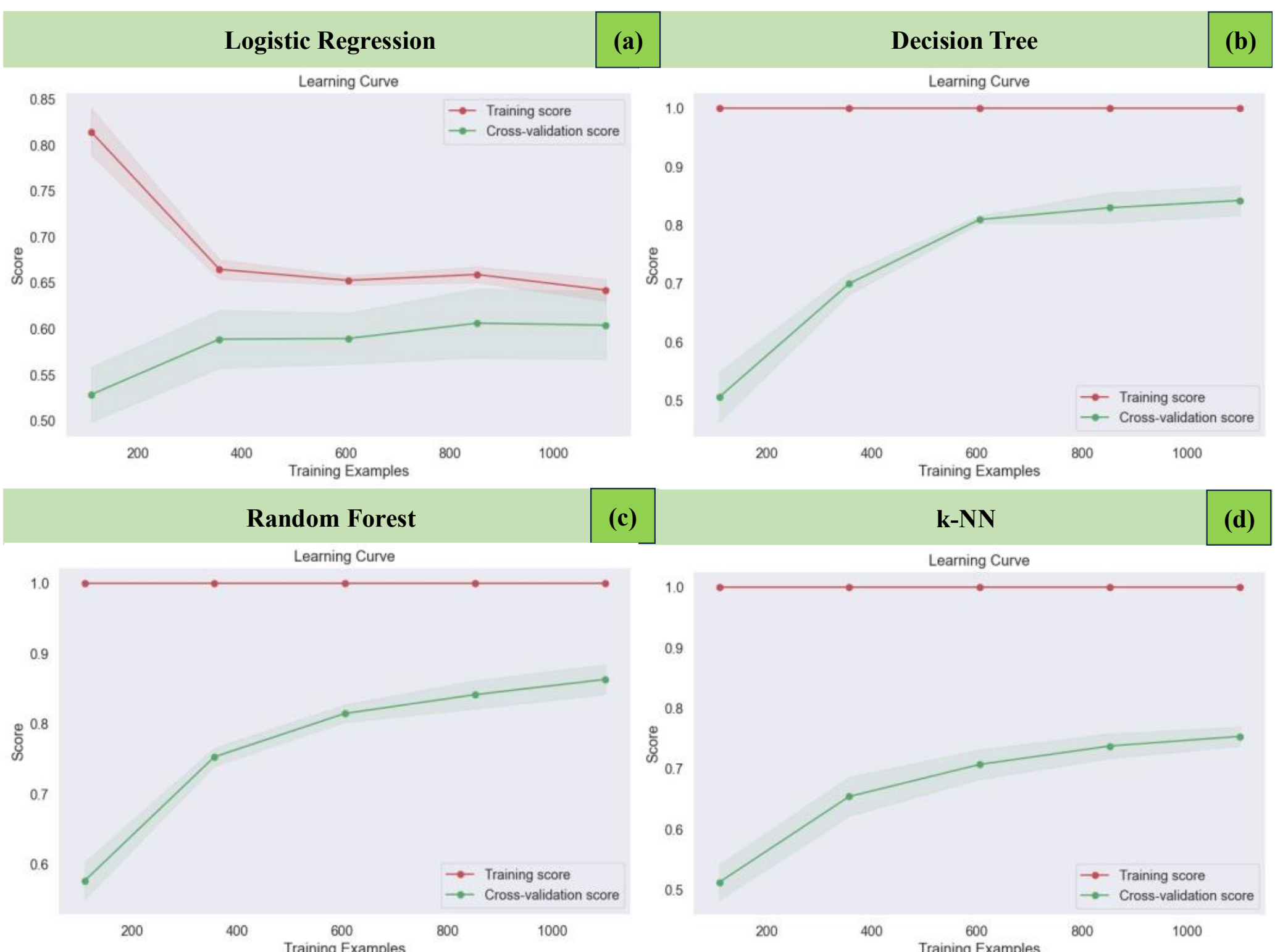

**Figure 14.** Learning Curve. (a) Logistic Regression, (b)Decision Tree, (c) Random Forest, and (d) k-NN Models.

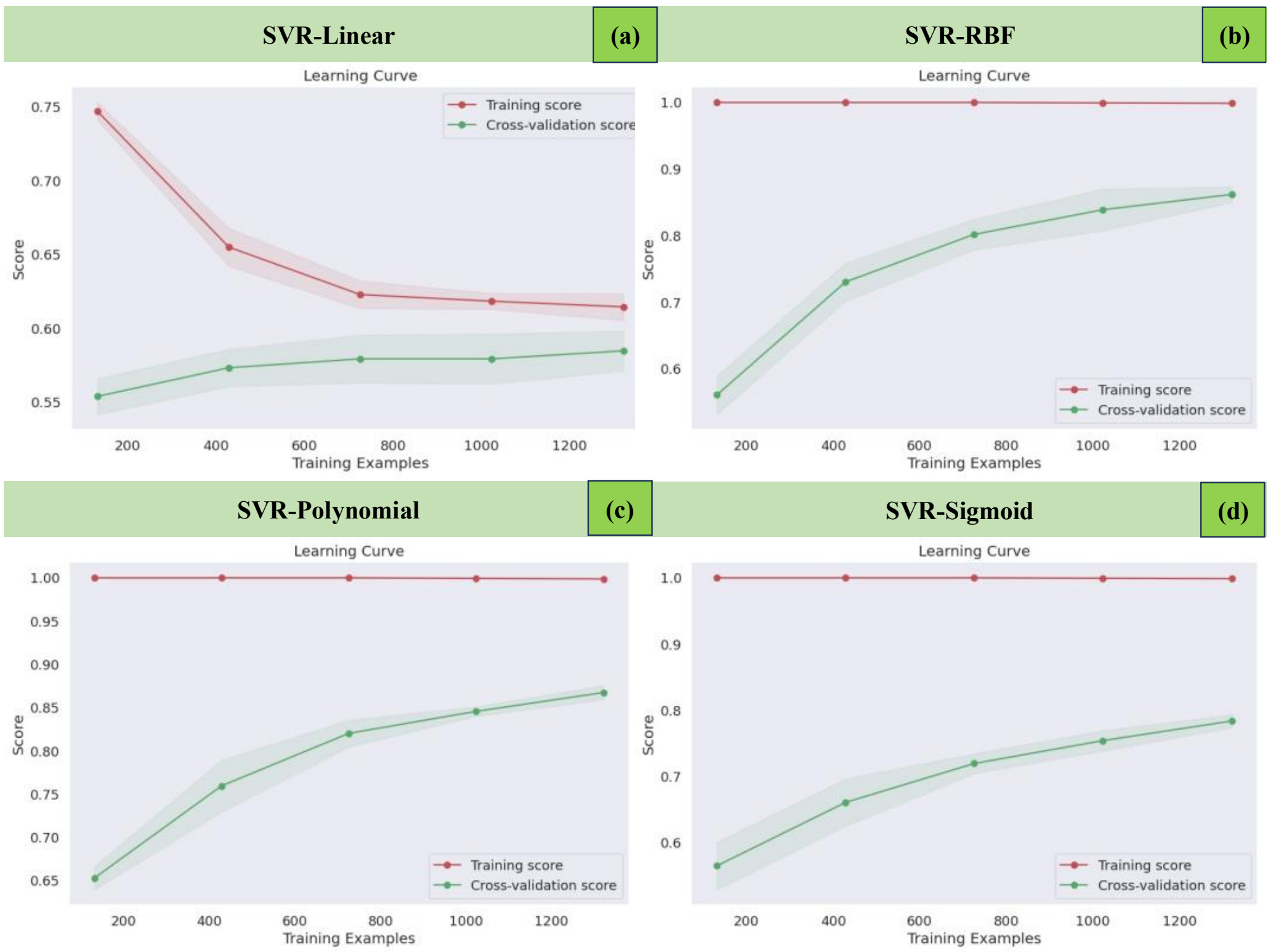

**Figure 15.** Learning Curve. (a) Linear, (b) RBF, (c) Polynomial, and (d) Sigmoid Kernels

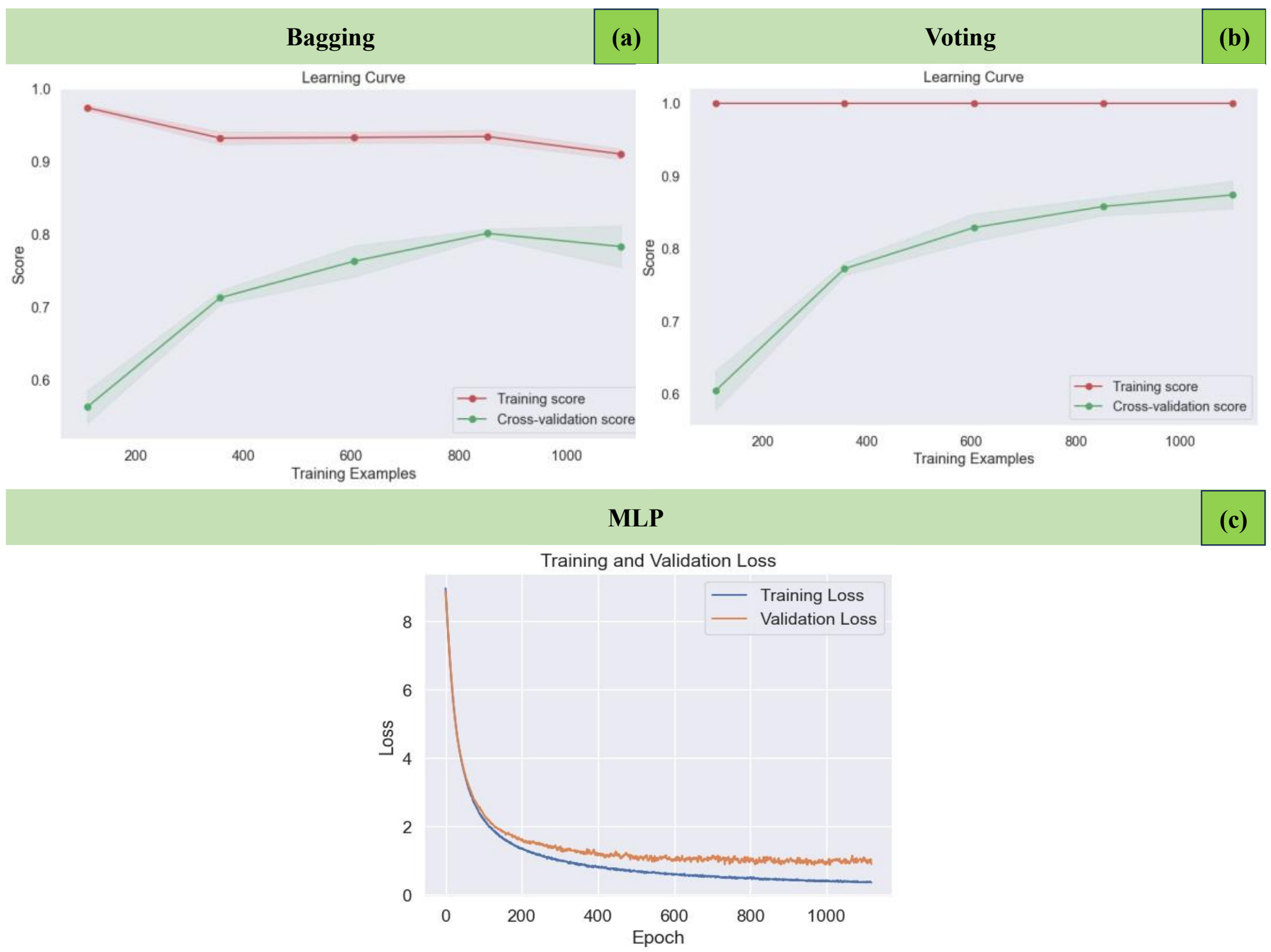

**Figure 16.** Learning Curve. (a) Bagging, (b) Voting, and (c) NN Models.

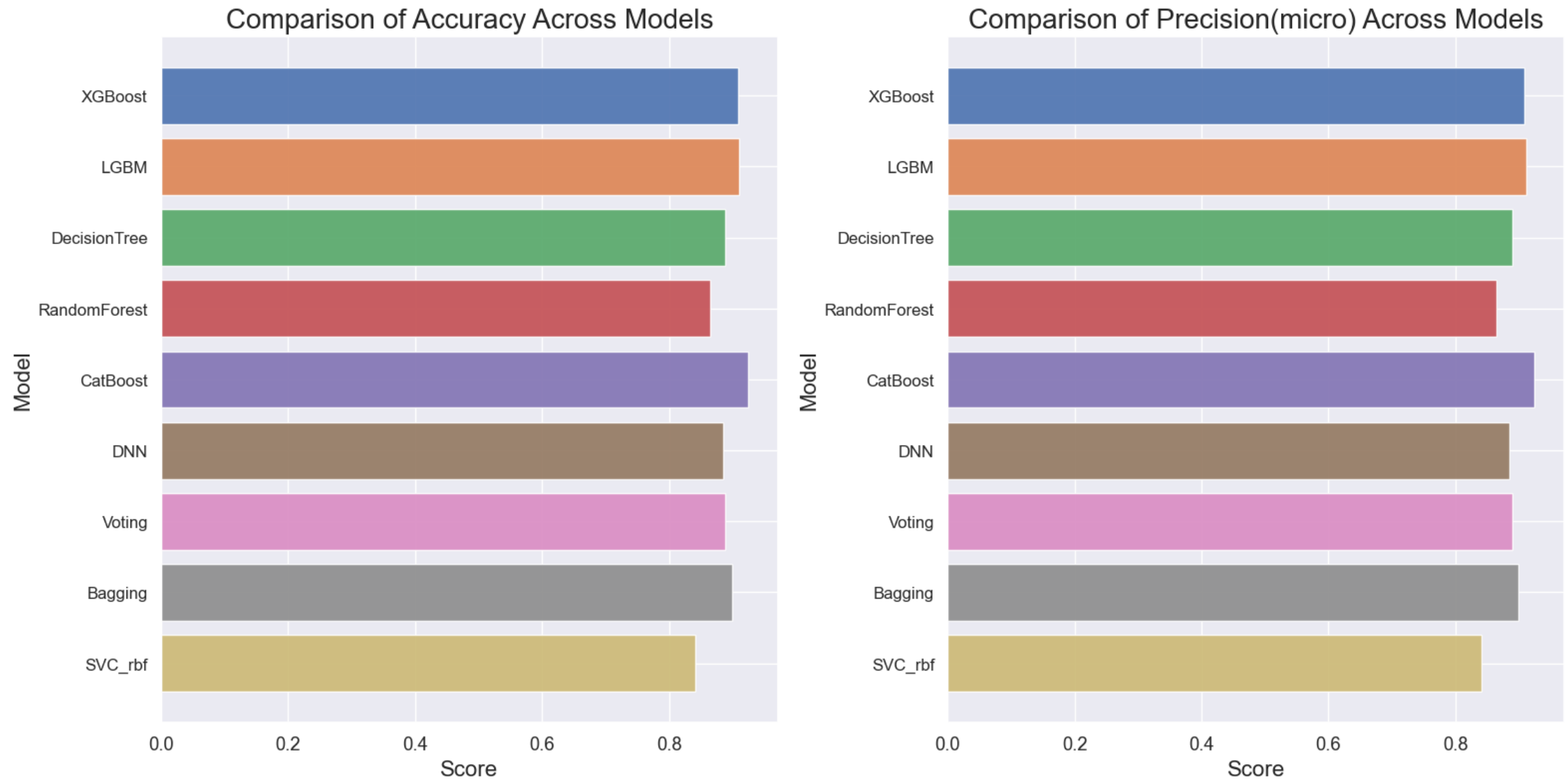

**Figure 17.** Comparing the top nine ML models in AM-DefectNet while considering various metrics.

**Table 3.** Recent works have utilized ML models in AM to yield promising results.

| Model | Target | Precision | Accuracy | F-1 Score | Recall | Reference |
|---|---|---|---|---|---|---|
| **SVM** | Porosity Detection | - | 89 | - | - | [17] |
| | Detect part discontinuities | - | 85 | - | - | [18] |
| | Underheating, medium underheating, normal, medium overheating, overheating | - | 89.13% | - | - | [16] |
| | Desirable, balling, severe keyholing, keyholing porosity, or under-melting | - | 85.1% | - | - | [19] |
| | Porosity | - | 89.36% | - | - | [17] |
| **Decision Tree** | Porosity Detection | - | 79 | - | - | [17] |
| **Linear Discriminant Analysis** | Porosity Detection | - | 82 | - | - | [17] |
| **K-NN** | Porosity Detection | - | 78 | - | - | [17] |
| | Keyhole, lack of fusion | - | 77% | - | - | [20] |
| | Porosity | - | 78% | - | - | [16] |
| **Ensemble** | Porosity Detection | - | 85 | - | - | [17] |
| **Neural Network (NN)** | Defect detection | - | 86 | - | - | [21] |
| | Porosity Detection | - | 84 | - | - | [17] |
| | Classifying different quality levels | - | 76–86 | - | - | [22] |
| | Classifying different parts complexity | - | 55–88 | - | - | [22] |
| | Defect size | 86.7 % | 82.9 % | 82.0 % | 77.8 % | [21] |
| | Balling, lack-of-fusion, conduction, key hole | - | 80% | - | - | [13] |

## 4. Conclusion

Additive Manufacturing is a sophisticated multi-physics process influenced by numerous process parameters and the thermal-affected melt pool zone. Defects such as keyhole formation, balling phenomenon, and lack of fusion (LOF) are common in AM-built products, with material properties playing a crucial role in their occurrence. In-situ and ex-situ monitoring techniques, along with numerical modeling, are commonly employed to identify defects in AM productions. However, AM monitoring presents significant challenges, particularly in terms of time consumption and effectiveness. To address these challenges, ML techniques offer a reliable and accurate solution. In our study, we introduced a novel benchmark named AM-DefectNet, leveraging 15 ML models to classify AM defects. We evaluated the performance of these models using four key metrics - accuracy, precision, recall, and F1-score - across three groups: macro, micro, and weighted. Our findings highlighted several significant results:

1. Among the 15 models considered in our benchmark, CatBoost emerged as the top-performing algorithm, achieving an accuracy of 92.47%. Followed closely were LGBM and XGBoost, with accuracies of 91.08% and 90.89%, respectively. Notably, the leading models primarily consisted of non-linear tree-based algorithms, with the Deep Neural Network (DNN) also displaying competitive performance.
2. CatBoost demonstrated superior performance in classification tasks, surpassing other gradient boost algorithms in terms of precision, recall, F1-score, and overall accuracy. The model exhibited robust performance across different classes, further validating its effectiveness in defect classification tasks.
3. Learning curves provided valuable insights into the potential for further performance improvement and the reasons behind suboptimal model performance. These curves depicted the evolution of model performance with increasing training data, offering insights into model fitting and data requirements.

In summary, our study underscores the efficacy of ML techniques, particularly CatBoost, in addressing the challenges of defect classification in AM. By establishing the AM-DefectNet benchmark and providing comprehensive insights into model performance, we contribute to the advancement of defect detection methodologies in additive manufacturing processes.